\documentclass[a4paper,10pt]{article}
\usepackage{fullpage}
\usepackage[numbers,sort&compress]{natbib}
\usepackage{authblk}

\usepackage[utf8]{inputenc}
\usepackage{amsmath}
\usepackage{amsfonts}
\usepackage{amssymb}
\usepackage{graphicx}
\usepackage{multirow}
\usepackage{array}
\usepackage{subfigure}
\usepackage{algorithmic}
\usepackage{algorithm}
\usepackage{color}
\usepackage{sidecap}
\usepackage{booktabs}

\definecolor{darkgreen}{rgb}{0,0.6,0.2}

\title{Investigating echo state networks dynamics by means of recurrence analysis}

\author[1]{Filippo Maria Bianchi\thanks{filippo.m.bianchi@uit.no}}
\author[2,3]{Lorenzo Livi\thanks{lorenzo.livi@usi.ch}\thanks{Corresponding author}}
\author[2,3]{Cesare Alippi\thanks{cesare.alippi@polimi.it}}
\affil[1]{Dept. of Physics and Technology, UiT the Arctic University of Norway, Hansine Hansens veg 18, 9019 Troms\o{}, Norway}
\affil[2]{Dept. of Electronics, Information, and Bioengineering, Politecnico di Milano, Piazza Leonardo da Vinci 32, 20133 Milano, Italy}
\affil[3]{ALaRI, Faculty of Informatics, Universit\`a della Svizzera Italiana, Via G. Buffi 13, 6904 Lugano, Switzerland}

\providecommand{\keywords}[1]{\textbf{\textit{Keywords---}} #1}

\begin{document}

%\title{Investigating echo state networks dynamics by means of recurrence analysis}

%\author{
%		Filippo Maria Bianchi,
%		Lorenzo Livi,~\IEEEmembership{Member,~IEEE},
%       	and Cesare Alippi,~\IEEEmembership{Fellow,~IEEE}
%\thanks{Manuscript received ; revised .}
%\thanks{Filippo Maria Bianchi is with the Dept. of Physics and Technology, UiT the Arctic University of Norway, Hansine Hansens veg 18, 9019 Troms\o{}, Norway (e-mail: filippo.m.bianchi@uit.no).}
%\thanks{Lorenzo Livi and Cesare Alippi are with the Dept. of Electronics, Information, and Bioengineering, Politecnico di Milano, Piazza Leonardo da Vinci 32, 20133 Milano, Italy and ALaRI, Faculty of Informatics, Universit\`a della Svizzera Italiana, Via G. Buffi 13, 6904 Lugano, Switzerland (e-mail: lorenzo.livi@usi.ch, cesare.alippi@polimi.it).}}

\maketitle

\begin{abstract}
In this paper, we elaborate over the well-known interpretability issue in echo state networks.
The idea is to investigate the dynamics of reservoir neurons with time-series analysis techniques taken from research on complex systems. Notably, we analyze time-series of neuron activations with Recurrence Plots (RPs) and Recurrence Quantification Analysis (RQA), which permit to visualize and characterize high-dimensional dynamical systems.
We show that this approach is useful in a number of ways. 
First, the two-dimensional representation offered by RPs provides a way for visualizing the high-dimensional dynamics of a reservoir. Our results suggest that, if the network is stable, reservoir and input denote similar line patterns in the respective RPs. Conversely, the more unstable the ESN, the more the RP of the reservoir presents instability patterns.
As a second result, we show that the $\mathrm{L_{max}}$ measure is highly correlated with the well-established maximal local Lyapunov exponent.
This suggests that complexity measures based on RP diagonal lines distribution provide a valuable tool to quantify the degree of network stability.
Finally, our analysis shows that all RQA measures fluctuate on the proximity of the so-called edge of stability, where an ESN typically achieves maximum computational capability.
We verify that the determination of the edge of stability provided by such RQA measures is more accurate than two well-known criteria based on the Jacobian matrix of the reservoir.
Therefore, we claim that RPs and RQA-based analyses can be used as valuable tools to design an effective network given a specific problem.
\\\keywords{Echo state network; Recurrence plot; Recurrence quantification analysis; Time-series; Dynamics.}
\end{abstract}
%\begin{IEEEkeywords}
%Echo state network; Recurrence plot; Recurrence quantification analysis; Time-series; Dynamics.
%\end{IEEEkeywords}

%%%%%%%%%% INTRODUCTION %%%%%%%%%%
\section{Introduction}

% Contextualization
Since the very first recurrent neural network architectures, attempts have been made to describe and understand the internal dynamics of the system -- see e.g. \cite{1000129} and references therein. Nowadays, such efforts found renewed interest by those researchers trying to ``open the black-box'' \cite{schiller2005analyzing,marichal2015analysis,sussillo2013opening,4435137}. This comes natural as recent advances in various fields, such as neurosciences and biophysical systems modeling, demand understanding of the inner mechanism that drives the inductive inference in order to produce novel scientific results \cite{sussillo2014neural}.

% RC and ESN
Reservoir computing is a class of state-space models characterized by a fixed state transition structure, the \textit{reservoir}, and a trainable, memory-less \textit{readout} layer \cite{verstraeten2007experimental,lukovsevivcius2009reservoir,dambre2012information}. A reservoir must be sufficiently complex to capture all salient features of the inputs, behaving as a time dependent, non-linear kernel function, which maps the inputs into a higher dimensional space.
The reservoir is typically generated with a pseudo-random procedure, not rarely driven by a set of rules-of-thumb and a trial-and-error approach.
A popular reservoir computing architecture is the Echo State Network (ESN) \cite{jaeger2002adaptive}, a recurrent neural network with a non-trainable, sparse recurrent reservoir and an adaptable (usually) linear readout mapping the reservoir to the output.
ESN reservoir characterization and design attracted significant research efforts in the last decade \cite{strauss2012design,gallicchio2011architectural,ma2014direct}.
This is mostly due to the puzzling behavior of the reservoir, which, although randomly initialized, has shown to be effective in modeling non-linear dynamical systems of various nature \cite{scardapane2015,bianchi2015prediction,liu2012data,7286732,6480841,Skowronski2007414}.
This also motivated researchers to better understand, and hence control, the dynamics of reservoirs given the fact they are driven by input signals.
The many existing approaches to reservoir design can be roughly partitioned as either supervised or unsupervised \cite{lukovsevivcius2009reservoir}.
In the former case, the output of the system is taken into account to analyze, and then accordingly design, the reservoir.
In the latter case, the reservoir is designed by considering algebraic/topological properties of the weight matrix \cite{xue2007decoupled,boccato2014self,6105577} or criteria based on statistics of the neuron activations \cite{ozturk2007analysis,schrauwen2008improving}.
Recently, it was shown that topologies designed according to deterministic criteria \cite{rodan2012simple,rodan2011minimum} produced state-of-the-art results in all major benchmarks, having a more contractive dynamics with respect to randomly generated reservoirs.
Another interesting reservoir design was proposed in \cite{appeltant2011information}. The authors showed that, for certain tasks, a large (randomly connected) reservoir could be replaced by a single non-linear neuron with delayed feedback.
It is worth citing that also the nature of the neuron activation function (spiking vs analog) has been studied in this context \cite{busing2010connectivity}, suggesting that networks operating with spiking neurons achieve inferior performance at the current state of research, possibly due to the adopted complex pre- and post-processing strategies.
Input-dependent measures to characterize the reservoir dynamics have been proposed in \cite{verstraeten2009quantification}, which considered the temporal profile of the Jacobian of the reservoir, rather than static quantities such as the spectral radius.
Another input-dependent method for describing the dynamic profile of the reservoir is proposed in \cite{boedecker2012information}. The authors demonstrated that both transfer entropy (the predictive information provided by a source about a destination that was not already contained in the destination history) and active information storage (the amount of past information that is relevant to predict the next state) calculated on the reservoir neurons are maximized right on the transition to an unstable regime.

% RP
The use of Poincar\'{e} recurrence of states provides fundamental information for the analysis of autonomous dynamical systems \cite{marwan2007recurrence}.
This follows from Poincar\'{e}'s theorem, which guarantees that the states of a dynamic system must recur during its evolution.
In other terms, the system trajectory in phase space must return in an arbitrarily small neighbourhood of any of the previously visited states with probability one.
Recurrences contain all relevant information regarding a system behavior in phase space and can be linked also with dynamical invariants (e.g., metric entropy) and features related to stability.
However, especially for high-dimensional complex systems, the recurrence time, which is the time elapsed between recurring states, is difficult to calculate even when assuming full analytical knowledge of the system.

Recurrence Plots (RPs) \cite{marwan2007recurrence,marwan2011avoid,eroglu2014entropy,marwan2013recurrence}, together with the computation of dynamical invariants and heuristic complexity measures called Recurrence Quantification Analysis (RQA), offer a simple yet effective tool to analyze such recurrences starting from a time-series derived from the system under analysis.
An RP is a visual representation of recurrence time, which also provides information about the duration of the recurrence by inspecting line patterns \cite{marwan2005line}.
RPs are constructed by considering a suitable distance in the phase (equivalently, state) space and a threshold used to determine the recurrence/similarity of states during the evolution of the system.
Recently, RPs have been extended to study heterogeneous recurrences \cite{yang2014heterogeneous}, i.e., qualitatively different types of recurrences in phase space. It is worth citing also the rapidly developing field of recurrence networks \cite{donner2010recurrence}, whose main goal is to exploit complex network methods to analyze time-series related to dynamical systems.

% Countribution
In this paper, we address the interpretability issue of ESNs by analyzing the dynamics of the reservoir neuron activations with RPs and RQA complexity measures.
To the best of our knowledge, recurrence analysis has never been used to investigate reservoirs or ESNs.
RPs and RQA-based techniques allow the designer to visualize and characterize (high-dimensional) dynamical systems starting from a matrix encoding the recurrences of the system states over time.
We show that RPs and RQA-based analyses allow to deduce important and consistent conclusions about the behavior of the network. Therefore, we suggest that they can be used as a valuable analysis tool to design a network for the problem at hand.
The novelty content of what proposed can be summarized as:
\begin{enumerate}
	\item We offer a method to visualize high-dimensional dynamics of input-driven ESNs through a two-dimensional representation encoding recurrences of states. Our results show that, if the network is stable, line patterns of RPs related to reservoir dynamics and those of the input signal are similar. This similarity is lost as soon as the network becomes unstable;
	\item We comment that the degree of instability of the network can be quantified by means of a measure derived from the RP. This measure is named $\mathrm{L_{max}}$ and is based on the diagonal lines of the RP only (hence it is computationally cost effective). To support this claim, we show that such a measure is correlated with the well-established maximal local Lyapunov exponent over the entire range of the control parameters;
	\item We show that RQA measures can help the application designer to tune important network parameters (e.g., spectral radius and input scaling), in order to push the ESN in a critical state, where computational capability (defined in terms of prediction performance and short-term memory capacity) is maximized. Such a critical state (called edge of stability) depends on the dynamics of the network which, in turn, is influenced by the input signal. Therefore, this cannot be achieved by considering only static measures, e.g., taking into account only the reservoir topology. As we will show here, the proposed method based on RQA measures is sensitive also to the nature of the input signal driving the network.
\end{enumerate}

% Paper structure
This paper is structured as follows.
In Sec. \ref{sec:esn_reservoir_dynamics} we introduce a typical ESN model and discuss some important figures of merit that have been proposed in the literature to characterize a reservoir.
In Sec. \ref{sec:rp_esn}, we present the contribution of our work, i.e., we link recurrence analysis with ESNs and discuss how this new tool can be used in practice to design a network exposing both stability and forecasting accuracy properties.
In Sec. \ref{sec:exp}, we introduce experiments to support our claims. 
Finally, Sec. \ref{sec:conclusions} offers concluding remarks and future directions.

%%%%%%%%%% ESN %%%%%%%%%%
\section{Dynamics of the ESN reservoir}
\label{sec:esn_reservoir_dynamics}

\begin{SCfigure}[][ht!]
    \centering
    \includegraphics[scale=.8, keepaspectratio]{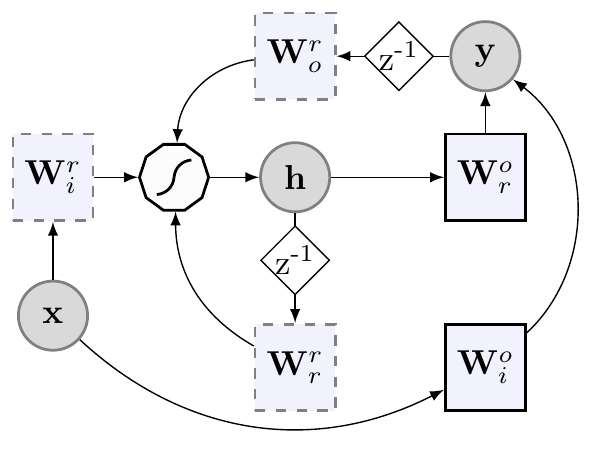}
    \caption{The ESN architecture. Circles represent input $\mathbf{x}$, state, $\mathbf{h}$, and output, $\mathbf{y}$, respectively. Solid squares $\mathbf{W}_{r}^{o}$ and $\mathbf{W}_{i}^{o}$, are the trainable matrices of the readout, while dashed squares, $\mathbf{W}_{r}^{r}$, $\mathbf{W}_{o}^{r}$, and $\mathbf{W}_{i}^{r}$, are randomly initialized matrices. The polygon represents the non-linear transformation performed by neurons and $\text{z}^{\text{-1}}$ is the unit delay.}
    \label{fig:esn}
\end{SCfigure}

A schematic representation of an ESN is shown in Fig. \ref{fig:esn}.
An ESN consists of a reservoir of $N_r$ nodes characterized by a non-linear transfer function $f(\cdot)$. At time instant $k$, the network is driven by the input $\mathbf{x}[k]\in \mathbb{R}^{N_i}$ and produces the output $\mathbf{y}[k] \in \mathbb{R}^{N_o}$, being $N_i$ and $N_o$ the dimensionalities of input and output, respectively. The weight matrices $\mathbf{W}_r^r \in \mathbb{R}^{N_r \times N_r}$ (reservoir internal connections), $\mathbf{W}_i^r \in \mathbb{R}^{N_i \times N_r}$ (input-to-reservoir connections), and $\mathbf{W}_o^r \in \mathbb{R}^{N_o \times N_r}$ (output-to-reservoir feedback connections) contain values in the $[-1, 1]$ interval drawn from a uniform distribution; a Gaussian distribution is another common choice.

An ESN is a discrete-time non-linear system with feedback, whose model reads:
\begin{align}
\label{eq:esn_state}
\mathbf{h}[k] & = f\left( \mathbf{W}_{r}^{r} \mathbf{h}[k-1] + \mathbf{W}_{i}^{r} \mathbf{x}[k] + \mathbf{W}_{o}^{r} \mathbf{y}[k-1] \right); \\
\label{eq:esn_output}
\mathbf{y}[k] & = g\left( \mathbf{W}_{r}^{o} \mathbf{h}[k] + \mathbf{W}_{i}^{o} \mathbf{x}[k] \right).
\end{align}

Activation functions $f(\cdot)$ and $g(\cdot)$, both applied component-wise, are typically implemented as a sigmoidal (\emph{tanh}) and identity function, respectively.
The output weight matrices $\mathbf{W}_r^o \in \mathbb{R}^{N_r \times N_o}$ and $\mathbf{W}_i^o \in \mathbb{R}^{N_i \times N_o}$, which connect, respectively, reservoir and input to the output, represent the readout of the network.
The standard training procedure for such matrices requires solving a straightforward regularized least-square problem \cite{jaeger2001echo}.

Even though the three weight matrices $\mathbf{W}_{r}^{r}$, $\mathbf{W}_{o}^{r}$, and $\mathbf{W}_{i}^{r}$ are generated randomly, they can be suitably designed and scaled to obtain desired properties.
For instance, $\mathbf{W}_{r}^{o}$ can scale with multiplicative constant $\omega_{o}$; in this work, we set $\omega_{o} = 0$, which has the effect of removing the output feedback connection.
$\mathbf{W}_{i}^{r}$ is controlled by scalar parameter $\omega_i$.
Since the gain of the sigmoid non-linearity in the neurons is largest around the origin, the scaling coefficient $\omega_{i}$ of $\mathbf{W}_{i}^{r}$ determines the amount of non-linearity introduced by the processing units. 
In particular, inputs far from zero tend to drive the activation of the neurons towards saturation where they show more non-linearity.
Finally, the spectral radius of $\mathbf{W}_r^r$, denoted as $\rho(\mathbf{W}_{r}^{r})$ or $\rho$, controls important properties as discussed in the sequel.

% ESP property
An ESN is typically designed so that the influence of past inputs on the network gradually fades away and the initial state of the reservoir is eventually washed out.
This is granted by the Echo State Property (ESP), which ensures that, given any input sequence taken from a compact set, future trajectories of any two different initial states become indistinguishable.
ESP was originally investigated in \cite{jaeger2001echo} and successively in \cite{yildiz2012re}; we refer the interested reader to \cite{manjunath2013echo} for a more recent definition, where also the influence of input is explicitly accounted for.
In ESNs with no output feedback, as in our case, the state update of equation (\ref{eq:esn_state}) reduces to
\begin{equation}
\label{eq:esn_state_nooutputfeedback}
\mathbf{h}[k] = f(\mathbf{W}_{r}^{r}\mathbf{h}[k-1] + \mathbf{W}_{i}^{r}\mathbf{x}[k]).
\end{equation}

In order to study the stability of the network, we analyze the Jacobian of the state update (\ref{eq:esn_state_nooutputfeedback}) of the reservoir and generate a derived measure, the maximal local Lyapunov exponent ($\lambda$). Such a quantity is used to approximate (for an autonomous system) the separation rate in phase space of trajectories having very similar initial conditions.
$\lambda$ was proposed to characterize a reservoir, and demonstrated its efficacy in designing a suitable network configuration in several applications \cite{verstraeten2006reservoir,verstraeten2009quantification,verstraeten2007experimental}.
$\lambda$ is calculated by considering the Jacobian at time $k$, which can be conveniently expressed if neurons are implemented with a \textit{tanh} activation function as
\begin{align}
\label{eq:jacob}
&\mathbf{J}(h[k]) = \\
&\nonumber\left[\begin{array}{cccc}
1 - (h_1[k])^2 & 0 & \ldots & 0 \\
0 & 1 - (h_2[k])^2 & \ldots & 0 \\
\vdots & \vdots  & \ddots & \vdots \\
0 & 0 & \ldots & 1 - (h_{N_r}[k])^2 \\
\end{array}\right]
\mathbf{W}_{r}^{r} \ ,
\end{align}
where $h_l[k]$ is the activation of the $l$-th neuron, with $l=1, 2, ..., N_r$.
$\lambda$ is then computed as
\begin{equation}
\label{eq:MLLE}
\lambda = \max_{n=1, ..., N_r} \frac{1}{K} \sum_{k=1}^{K} \log \left( r_n[k] \right),
\end{equation}
where $r_n[k]$ is the module of $n$-th eigenvalue of $\mathbf{J}(h[k])$ and $K$ is the total number of time-steps in the considered trajectory.
Typically, the stable--unstable transition is detected numerically by considering the sign of $\lambda$ (\ref{eq:MLLE}).
In autonomous systems, $\lambda>0$ indicates that the dynamics is chaotic.

Local, first-order approximations provided by Eq. \ref{eq:jacob} are useful also for studying the stability of a (simplified) reservoir operating around the zero state, $\mathbf{0}$.
In fact, implementing $f(\cdot)$ as a \textit{tanh} assures $f(\mathbf{0})=\mathbf{0}$, i.e., $\mathbf{0}$ is a fixed point of the ESN dynamics.
Therefore, by linearizing (\ref{eq:esn_state_nooutputfeedback}) around $\mathbf{0}$ and assuming a zero-input, we obtain from (\ref{eq:jacob})
\begin{equation}
\label{eq:lin_dyn_state}
\mathbf{h}[k] = \mathbf{J}(\mathbf{0})\mathbf{h}[k-1] = \mathbf{W}_{r}^{r}\mathbf{h}[k-1].
\end{equation}

Linear stability analysis of (\ref{eq:lin_dyn_state}) suggests that, if $\rho(\mathbf{W}_{r}^{r}) < 1$, then the dynamic around $\mathbf{0}$ is stable.
In the more general case, the non-linearity of the sigmoid functions in (\ref{eq:esn_state_nooutputfeedback}) forces the norm of the state vector of the reservoir to remain bounded.
Therefore, the condition $\rho(\mathbf{W}_{r}^{r}) < 1$ looses its significance and does not guarantee stability when the system deviates from a small region around $\mathbf{0}$ \cite{verstraeten2006reservoir}.
This means that, actually, it is possible to find reservoirs (\ref{eq:esn_state_nooutputfeedback}) having $\rho(\mathbf{W}_{r}^{r}) > 1$ while still possessing the ESP with stability.
In fact, the effective local gain decreases when the operating point of the neurons shifts toward the positive/negative branch of the sigmoid, where stabilizing saturation effects start to influence the excitability of reservoir dynamics \cite{yildiz2012re}.
In the more realistic and useful scenario where the input driving the network is a generic (non-zero) signal, a sufficient condition for the ESP is met if $\mathbf{W}_{r}^{r}$ is diagonally Schur-stable, i.e., if there exists a positive definite diagonal matrix, $\mathbf{P}$, such that $(\mathbf{W}_{r}^{r})^{T} \mathbf{P} \mathbf{W}_{r}^{r} - \mathbf{P}$ is negative definite \cite{yildiz2012re}.
However, this recipe is fairly restrictive in practice. In many cases, such a sufficient condition might generate reservoirs that are not rich enough in terms of provided dynamics, since the use of a conservative scaling factor might compromise the amount of memory in the network and thus the ability to accurately model a given problem.
Therefore, for most practical purposes, the necessary condition $\rho(\mathbf{W}_{r}^{r}) < 1$ is considered ``sufficient in practice'', because if the spectral radius is less than 1, the state update map is contractive with high probability, regardless of the input and given a sufficiently large reservoir \cite{6105577}.

% Edge of stability/computational power/memory capacity - rho close to 1
The number of reservoir neurons and the bounds on $\rho$ can be used for a na\"{\i}ve quantification of the computational capability of a reservoir \cite{yildiz2012re}.
However, those are static measures that only consider the algebraic properties of $\mathbf{W}_{r}^{r}$, without taking into account other factors, such as the input scaling $\omega_{i}$ and the particular properties of the given input signals.
Moreover, it is still not clear how, in a mathematical sense, these stability bounds relate to the actual ESN dynamics when processing non-trivial input signals \cite{manjunath2013echo}. 
In this context, the idea of pushing the system toward the so-called ``edge of stability'' (also called edge of criticality) has been explored. \cite{langton1990computation, bertschinger2004real,legenstein2007edge} show that several dynamical systems, among which randomly connected recurrent neural networks, achieved the highest computational capabilities when moving toward the unstable (sometime even chaotic) regime, where the ESP is lost and the system enters into an oscillatory behavior.
This justifies the use of spectral radii above the unity in some practical applications.
It is important to remark that, although the critical state might maximize the computational capability, there are also tasks that either require very little in terms of computation or reservoirs with fast reaction times and fast locking into attractor states (e.g., multistable switching circuits \cite{jaeger2001short}).

Further descriptors used for characterizing the dynamics of a reservoir are based on information-theoretic quantities, such as the (average) transfer entropy and the active information storage one \cite{boedecker2012information}. The authors have shown that such quantities peak right when $\lambda>0$.
In addition, also the minimal singular value of the Jacobian (\ref{eq:jacob}), denoted as $\eta$, was demonstrated to be an accurate predictor of ESN performance, providing more accurate information regarding the ESN dynamics than both $\lambda$ and $\rho$ \cite{verstraeten2009quantification}.
Finding hyper-parameters that maximize $\eta$ results in a dynamical system that is far from singularity, it has many degrees of freedom, a good excitability, and it separates well the input signals in phase space \cite{verstraeten2009quantification}.

Complementary to $\lambda$, $\eta$, and the aforementioned information-theoretic descriptors, in this study we propose to use RPs and the related RQA measures for characterizing the dynamics of an ESN when driven by an input signal.

%%%%%%%%%% RP and RQA %%%%%%%%%%
\section{Analyzing ESN dynamics by investigating recurrences of neuron activations}
\label{sec:rp_esn}

In this section, we discuss how the input-driven dynamics of an ESN reservoir (\ref{eq:esn_state_nooutputfeedback}) can be fruitfully analyzed by means of RP-based techniques.

\begin{figure}[ht!]
\centering
\includegraphics[width=\columnwidth]{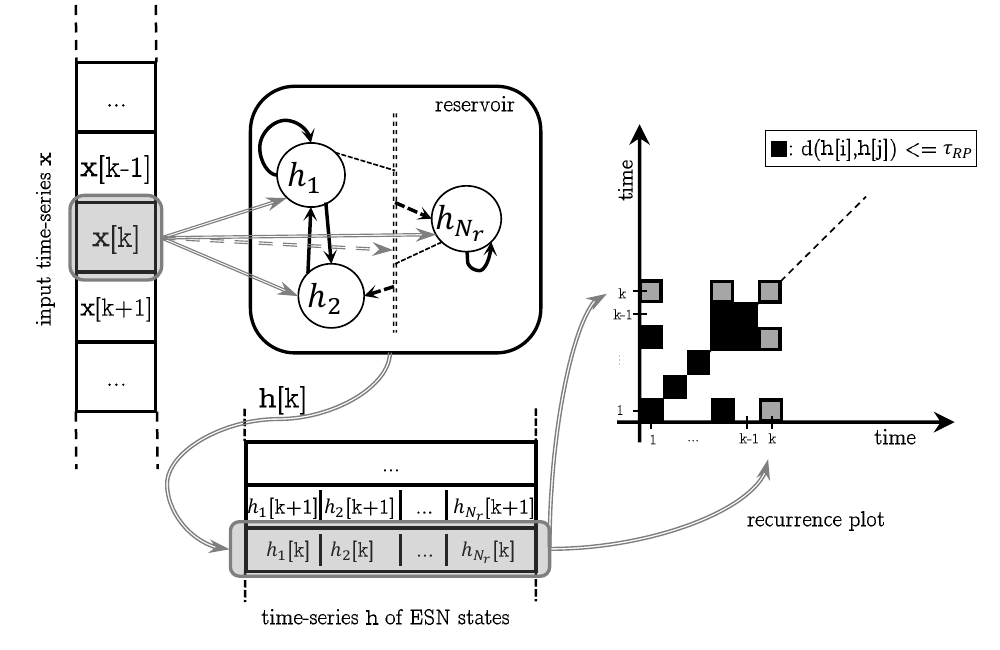}
\caption{When $\mathbf{x}[k]$ is fed as input to the $N_r$ neurons of the ESN reservoir, the internal state is updated to $\mathbf{h}[k] = [ h_1[k], h_2[k], \ldots ,h_{N_r}[k] ]^T$, where $h_n[k]$ is the output of the $n$-th neuron. Once the time-series $\mathbf{h}$ is generated, the RP is constructed by using a threshold $\tau_{\mathrm{RP}}$ and a dissimilarity measure $d(\cdot, \cdot)$. If $d(\mathbf{h}[k], \mathbf{h}[i]) \leq \tau_{\mathrm{RP}}$, the cell of the RP in position $(k,i)$ is colored in black, otherwise it is left white. The elements in gray highlight the operations performed at time-step $k$.}
\label{fig:RP_build}
\end{figure}

The sequence of ESN states can be seen as a multi-variate time-series $\mathbf{h}$, consisting of $N_r$ state variables coming from the reservoir neuron activations.
An RP is constructed by calculating a $K\times K$ binary matrix $\mathbf{R}$. The generic element $R_{ij}$ is defined as
\begin{equation}
\label{eq:RM}
R_{ij} = \Theta(\tau_{\mathrm{RP}} - d(\mathbf{h}[i], \mathbf{h}[j])), \ \ 1\leq i,j \leq K,
\end{equation}
where $K$ is the length of the time-series, $d(\cdot, \cdot)$ is a dissimilarity measure operating in phase space, and $\Theta(\cdot)$ the Heaviside function: $\Theta(x)=0$ if $x<0$; $\Theta(x)=1$ otherwise.
$\tau_{\mathrm{RP}}>0$ is a user-defined threshold used to identify recurrences.
$\tau_{\mathrm{RP}}$ can be defined in different ways, but typically it represents a percentage of the average or the maximum phase space distance between the states.
The selection of an ``optimal'' threshold $\tau_{\mathrm{RP}}$ is a problem-dependent issue.
However, we comment that, usually, $\tau_{\mathrm{RP}}$ has an impact only on the quantitative information derived from an RP but does not affect the general properties of the system as seen through a recurrence analysis.
$\mathbf{R}$ is constructed by considering also a nonnegative dissimilarity measure $d(\cdot, \cdot)$, which evaluates the distance between states.
Typical examples include the Euclidean, Manhattan, or max-norm distances.
Several options have been exploited in the literature for $\tau_{\mathrm{RP}}$ and $d(\cdot, \cdot)$; we refer the reader to \cite{marwan2007recurrence} for further detailed discussions.
Fig. \ref{fig:RP_build} depicts the algorithmic steps required to generate an RP on ESN states.

Depending on the properties of the analyzed time-series, different line patterns emerge in an RP \cite{marwan2005line}.
Besides providing an immediate visualization of the system properties, starting from $\mathbf{R}$, it is possible to compute also several complexity measures, those associated with an RQA.
Such measures are defined by the distribution of both vertical/horizontal and diagonal line structures present in the RP and provide a numerical characterization of the properties associated with the underlying dynamics.

In Sec. \ref{sec:rqa}, we provide details about the RQA measures considered in this study.
In Sec. \ref{sec:design_stable_effective_net}, we discuss how a network designer should take into account the two important issues of stability and computational capability by using the tools here introduced.
Finally, in Sec. \ref{sec:visual_id} we explain how to interpret the RP and related RQA measures relative to ESN state sequences driven by different classes of inputs.

\subsection{RQA complexity measures}
\label{sec:rqa}

We assume here to have generated $\mathbf{R}$ according to (\ref{eq:RM}) on a time-series of length $K$.
Many of the following measures are based on the histograms $P(l)$ and $P(v)$, counting, respectively, the number of diagonal and vertical lines having specific lengths $l$ and $v$,
\begin{align}
\nonumber P(l)&=\sum_{i,j=1}^{K-l} (1-R_{i-1,j-1})(1-R_{i+l,j+l})\prod_{k=0}^{l-1} R_{i+k,j+k}; \\
\nonumber P(v)&=\sum_{i,j=1}^{K-v} (1-R_{i,j})(1-R_{i,j+v})\prod_{k=0}^{v-1} R_{i,j+k}. 
\end{align}
Such RQA measures cover different aspects of the dynamics (e.g., features related to the stability and complexity of the system evolution).
Abbreviations and notation are kept consistent with \cite{marwan2007recurrence}.\\

\textbf{RR.}
The recurrence rate is a measure of the density of recurrences in $\mathbf{R}$,
\begin{equation}
\label{eq:RR}
\mathrm{RR}=\frac{1}{K^2}\sum_{i,j=1}^{K} R_{ij}.
\end{equation}
Given Eq. \ref{eq:RM}, RR corresponds to the correlation sum, an important concept used in chaos theory.
RR could help also with the selection of $\tau_{\mathrm{RP}}$ when performing multiple tests on different conditions, e.g., by preserving the rate.\\

\textbf{DET.}
A measure of the determinism level of the system, based on the percentage of diagonal lines of minimum length $l_{\mathrm{min}}$,
\begin{equation}
\label{eq:DET}
\mathrm{DET}=\frac{\sum_{l=l_{\mathrm{min}}}^{K} lP(l)}{\sum_{l=1}^{K} lP(l)} \in[0, 1].
\end{equation}
A periodic system would have $\mathrm{DET}$ close to unity and close to zero for a signal with no time-dependency.
The threshold $l_{\mathrm{min}}$ can be used also to discriminate different forms of determinism, such as periodic and chaotic motions. In fact, it is known that in chaotic dynamics diagonal lines are very short -- due to exponential divergence in phase space.\\

$\textbf{L}_{\textbf{max}}.$
Maximum diagonal line length,
\begin{equation}
\label{eq:lmax}
\mathrm{L_{max}}=\max\{l_i\}_{i=1}^{N_l},\ \ N_l=\sum_{l\geq l_{\mathrm{min}}} P(l),
\end{equation}
where $l_i$ is $i$th diagonal line length, $N_l$ is the total number of diagonal lines, and $1\leq\mathrm{L_{max}}\leq\sqrt{2}K$.
This measure is related to the mean exponential divergence in phase space.
A measure of divergence can be obtained as
\begin{equation}
\label{eq:div}
\mathrm{DIV}=1/\mathrm{L_{max}}\in(0, 1].
\end{equation}
Chaotic systems do not present long diagonal lines, since trajectories diverge exponentially fast. As formally discussed in \cite{marwan2007recurrence}, it is possible to relate $\mathrm{L_{max}}$ with the correlation entropy of the system -- that is a dynamical invariant measure. Notably, correlation entropy can be estimated by using the diagonal lines distribution.
The correlation entropy is a lower-bound for the sum of the positive Lyapunov exponents, providing thus a formal connection with the analysis of sensitivity on initial conditions.\\

\textbf{LAM.}
Laminarity is a descriptor of the presence of laminar phases in the system,
\begin{equation}
\label{eq:lam}
\mathrm{LAM}=\frac{\sum_{v=v_{\mathrm{min}}}^{K} vP(v)}{\sum_{v=1}^{K} vP(v)}\in[0, 1],
\end{equation}
where $v_{\mathrm{min}}$ is a threshold for the minimal vertical line length to be considered.
Laminar phases denote states of the system that do not change or change very slowly for a number of consecutive time-steps.\\

\textbf{ENTR.}
A complexity measure of an RP with respect to the diagonal lines distribution,
\begin{equation}
\label{eq:ENTR}
\mathrm{ENTR}=-\sum_{l=l_{\mathrm{min}}}^{K} p(l) \ln(p(l)),
\end{equation}
where $p(l)=P(l)/N_l$. As usual, $0\leq\mathrm{ENTR}\leq\log(K)$.
Signals with no time-dependence present $\mathrm{ENTR}\simeq 0$, i.e., the diagonal lines distribution is fully concentrated on very short lines (e.g., single dots).
Conversely, such a measure assumes high values when the diagonal lines distribution become heterogeneous.\\

\textbf{SWRP.}
Entropy of a weighted RP. Such a measure has been recently introduced in \cite{eroglu2014entropy} and provides an alternative to ENTR (\ref{eq:ENTR}). It quantifies the complexity of scalar distributions of \textit{strengths} for each time-step, providing a solution for the border effects that might appear in the computation of ENTR.
The computation of SWRP considers the similarity matrix $\mathbf{\tilde{S}}$, whose elements are defined as $\tilde{S}_{ij} = \exp(-d(\mathbf{h}[i], \mathbf{h}[j]))$, where $d(\cdot, \cdot)$ is the same as in Eq. (\ref{eq:RM}).
The results do not depend on $\tau_{\mathrm{RP}}$ and are not influenced by the length $K$ of the time-series.
The strengths are used to measure the heterogeneity of the density of a given point in phase space.
The $i$-th strength is defined as: $s_i = \sum_{j=1}^K \tilde{S}_{ij}$.
The distribution of the strengths is represented with histograms using a user-defined number $B$ of bins.
SWRP measure is then defined as
\begin{equation}
\label{eq:swrp}
\mathrm{SWRP} = - \sum_{b=1}^B p(b) \ln p(b),
\end{equation}
where $P(b)$ is the number of elements in the $b$-th bin and $p(b) = P(b)/K$ is the respective probability; as usual, $0\leq\mathrm{SWRP}\leq\log(B)$.

\subsection{Design of a stable and effective network}
\label{sec:design_stable_effective_net}

Recurrence analysis can be used to (i) guarantee stability for a given configuration and (ii) tune the network, so that the highest computational capability can be achieved.
These aspects are treated in the sequel.\\

\noindent\textit{\textbf{Stability issue}}\\
Constructing an RP from $\mathbf{h}$ allows to investigate the network stability issue.
As mentioned before, the degree of stability can be calculated with $\mathrm{L_\mathrm{max}}$ (\ref{eq:lmax}): the higher $\mathrm{L_\mathrm{max}}$, the more stable the system.
In addition to $\mathrm{L_\mathrm{max}}$, the designer can have an immediate visual interpretation of stability by inspecting RP: short and erratic diagonal lines in RPs denote instability/chaoticity, while long diagonal lines denote regularity (e.g., a periodic motion). We show in the experimental section that $\mathrm{L_\mathrm{max}}$ is anticorrelated with $\lambda$ and hence it can be considered as a reliable indicator for the (input-dependent) degree of network stability.
In the experiments we highlight also that, if the network is stable, the dynamics of the input and reservoir produce very similar line patterns in the respective RPs.
If the network is unstable, then the aforementioned similarity is lost.
Therefore, RPs can be used by the designer as visual tools to analyze the response of the network to a specific input.\\

\begin{figure}[ht!]
\centering

    \subfigure[]{
    \includegraphics[width=0.49\columnwidth]{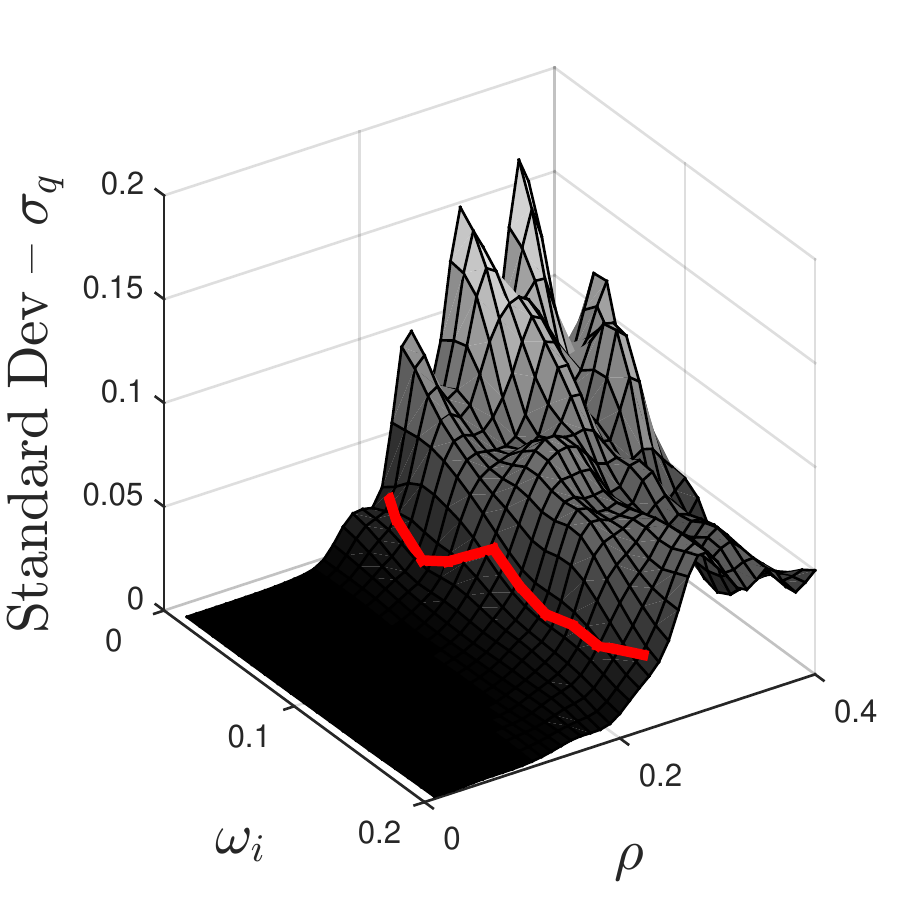}
    \label{fig:sin_fluct}}\hspace{-1.6em}%
    ~
		\subfigure[]{
    \includegraphics[width=0.49\columnwidth]{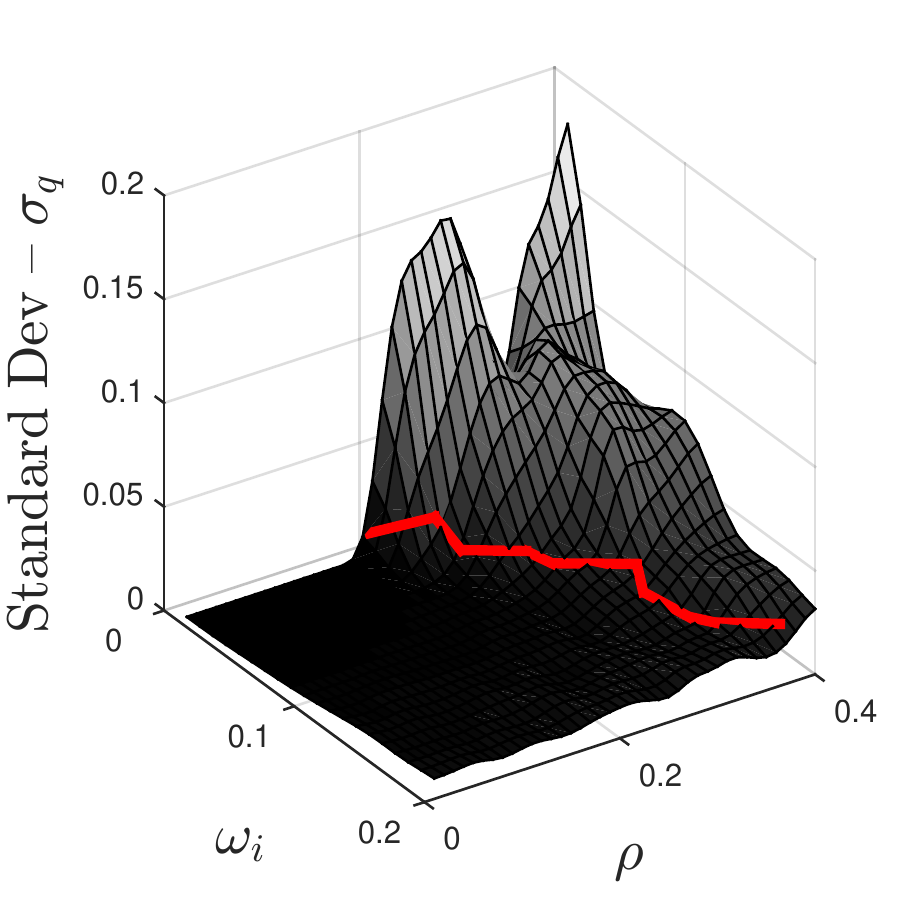}
    \label{fig:mg_fluct}}
		
\caption{Standard deviations of a RQA measure (RR) for different values of $\rho$ and $\omega_i$, relative to two different input signals. The edge of stability is identified in correspondence of the parameters configurations for which the standard deviation increases abruptly (shown as a red line), assuming values grater than its mean value.}
\label{fig:std_flucts}
\end{figure}

\noindent\textit{\textbf{Computational capability issue}}\\
A widespread criterion to determine the (signal-dependent) edge of stability relies on the sign of $\lambda$, which becomes positive when the system enters into a (globally) unstable regime.
We propose, instead, to use information derived from RR (\ref{eq:RR}), DET (\ref{eq:DET}), LAM (\ref{eq:lam}), ENTR (\ref{eq:ENTR}), and SWRP (\ref{eq:swrp}) for this purpose. We observed that, when such indexes start to fluctuate, the network achieves high prediction accuracy. This provides a guideline to the network designer for setting critical parameters, such as $\rho$ and $\omega_i$, in presence of input.
RQA measures, when evaluated on ESNs initialized with different weight matrices, assume very different values given $\rho$ and $\omega_i$ when the network in unstable.
Accordingly, it emerges that the standard deviation for such RQA measures is very high in such unstable regimes. We propose to identify the configurations of $\rho$ and $\omega_i$ which bring the system to the edge of stability with those for which we observe a sudden increment in the standard deviation.
Let $P$ and $\Omega$ be the domains for $\rho$ and $\omega_i$, respectively.
In this study, we calculate such configurations by checking when the standard deviations of RQA measures assume values grater than their mean values computed over $P\times\Omega$. 
For a given RQA measure, say $q$, we identify the edge of stability as a set of pairs $\{ (\rho^{1}, \omega_i^{1} ), (\rho^{2}, \omega_i^{2}), ..., (\rho^{|\Omega|}, \omega_i^{|\Omega|}) \} \subset P \times \Omega$.
For every $\omega_i^j \in \Omega$, we select the largest $\rho^j\in P$ such that
\begin{equation}
\label{eq:edge_criterion}
\sigma_q(\rho^j, \omega_i^j) \leq \bar{\sigma}_q,
\end{equation}
is satisfied. $\sigma_q(\rho^j, \omega_i^j)$ is the standard deviation of $q$ for the configuration $(\rho^j, \omega_i^j)$ and $\bar{\sigma}_q$ is the average value of the standard deviation computed over $P \times \Omega$.
Fig. \ref{fig:std_flucts} shows an illustrative example.
We stress that other criteria based on the fluctuations of RQA measures could be conceived.
However, as we will show in the experiments, such a criterion offers a more accurate description for the edge of stability than $\lambda$.

% MG RP

\subsection{Visualization and classification of reservoir dynamics}
\label{sec:visual_id}

In the following, we show how RPs permit to visualize, and hence classify, the reservoir dynamics when fed with inputs possessing important characteristics.
In doing so, we assume given a stable ESN described by (\ref{eq:esn_state_nooutputfeedback}); RPs are constructed following the procedure depicted in Fig. \ref{fig:RP_build}.
Although many classes of important signals/systems exist (with related sub-classes) \cite{marwan2007recurrence}, here we focus on the ability to discriminate between important classes for the input signals: (i) with/without time-dependence; (ii) periodic/non-periodic motions; (iii) laminar behaviours; (iv) chaotic dynamics; finally, (v) non-stationary processes.
%
% RP EXAMPLES
\begin{figure}[htp!]
\centering
	\subfigure[Gaussian white noise; $\tau_{\mathrm{RP}} = 0.4$]{
	\includegraphics[width=0.4\textwidth]{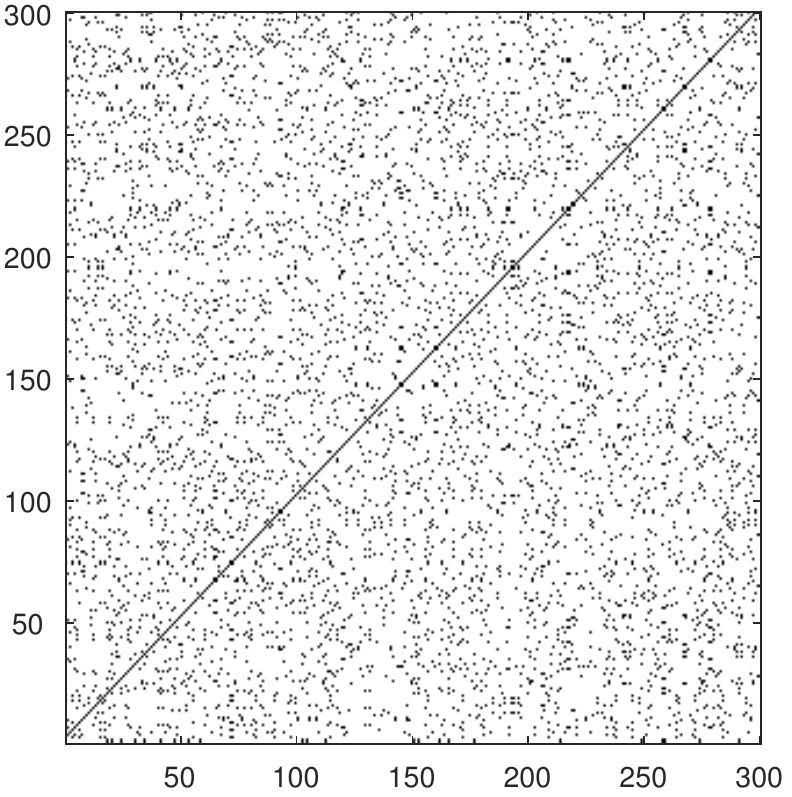}
	\label{fig:WN_SR01_T04}}\hspace{-1.2em}
	~
	\subfigure[Periodic; $\tau_{\mathrm{RP}} = 0.2$]{
	\includegraphics[width=0.4\textwidth]{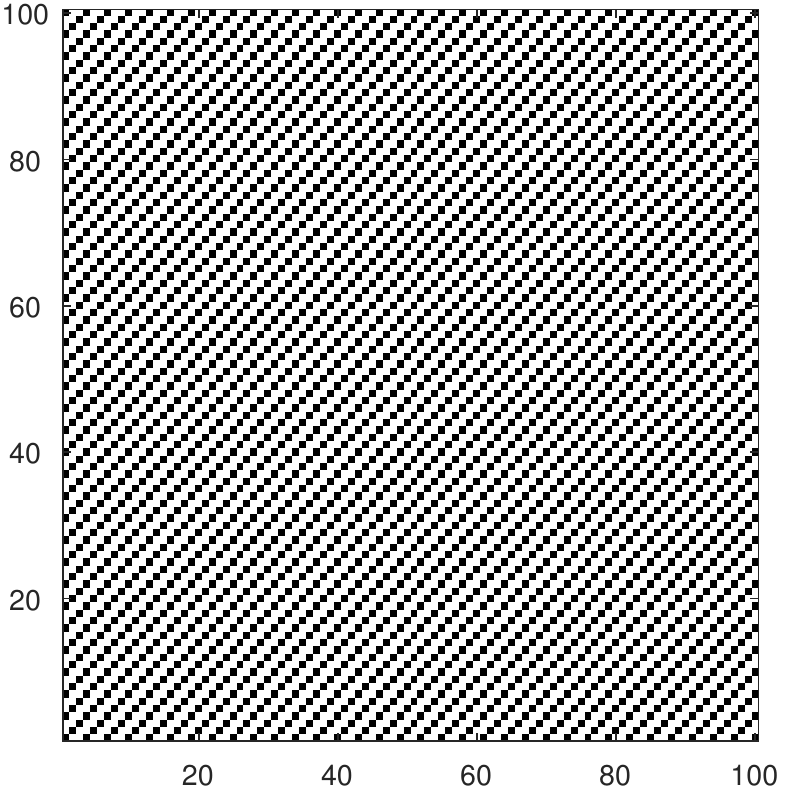}
	\label{fig:LM3.83_SR09_T03}}
	
    \subfigure[LM: laminar states; $\tau_{\mathrm{RP}} = 0.5$]{
	\includegraphics[width=0.4\textwidth]{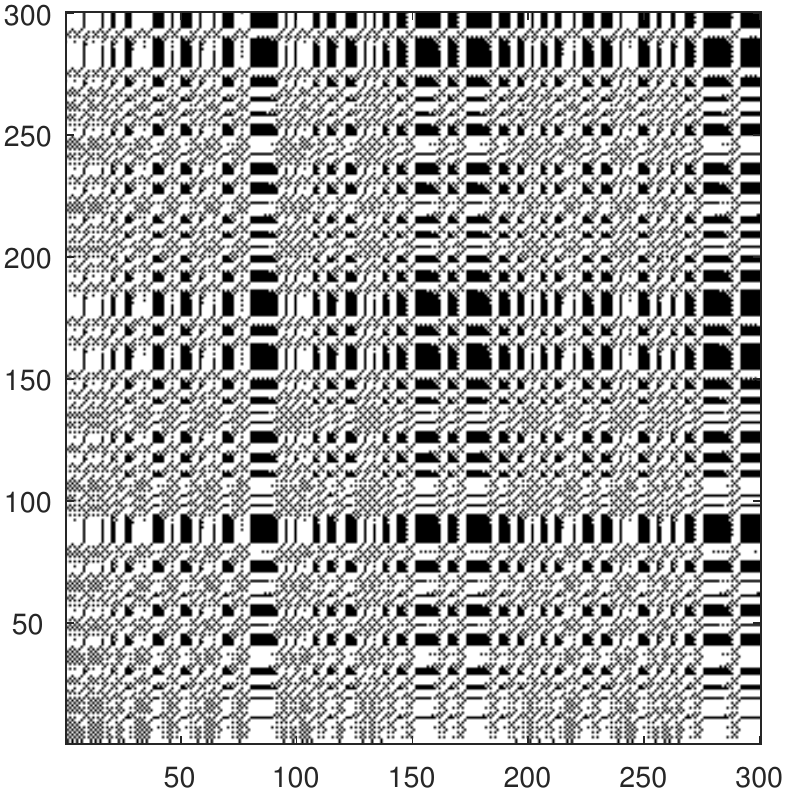}
	\label{fig:LM3.679_SR09_T05}}\hspace{-1.2em}
	~  		
	\subfigure[LM: chaos; $\tau_{\mathrm{RP}} = 0.2$]{
	\includegraphics[width=0.4\textwidth]{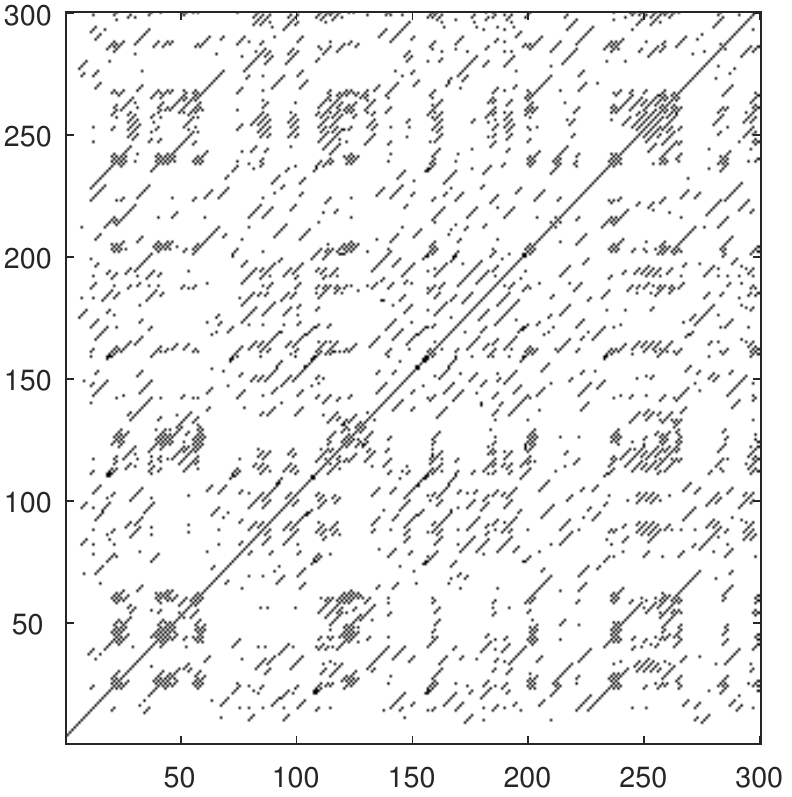}
	\label{fig:LM3.99_SR09_T03}}
	
	\subfigure[Brownian motion; $\tau_{\mathrm{RP}} = 0.2$]{
    \includegraphics[width=0.4\textwidth]{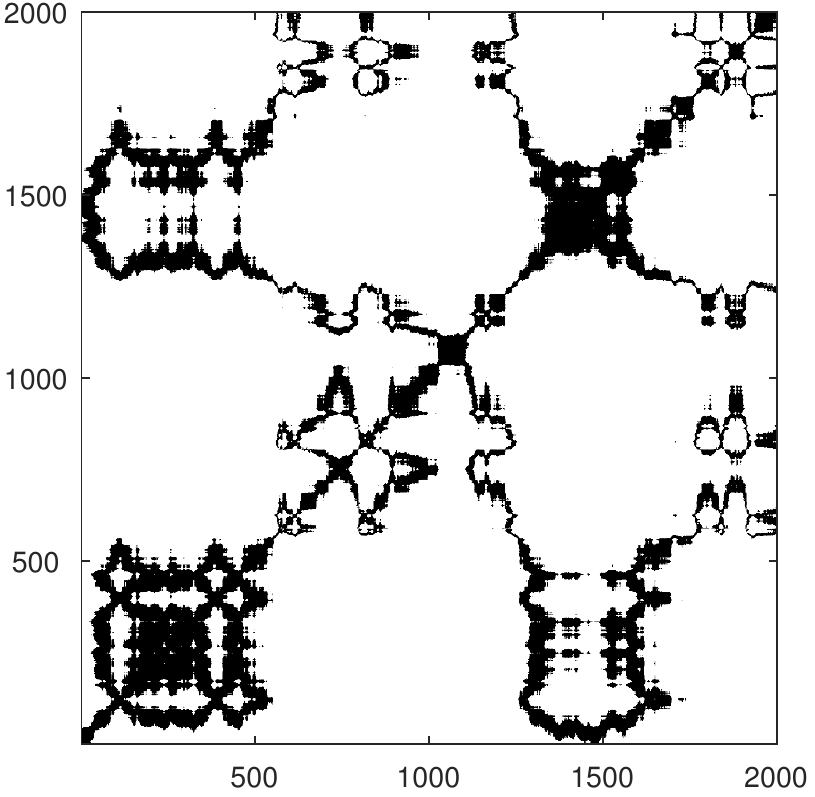}
    \label{fig:fBmP_SR0.9_T02}}
    ~
	\subfigure[Drift; $\tau_{\mathrm{RP}} = 0.2$]{
    \includegraphics[width=0.4\textwidth]{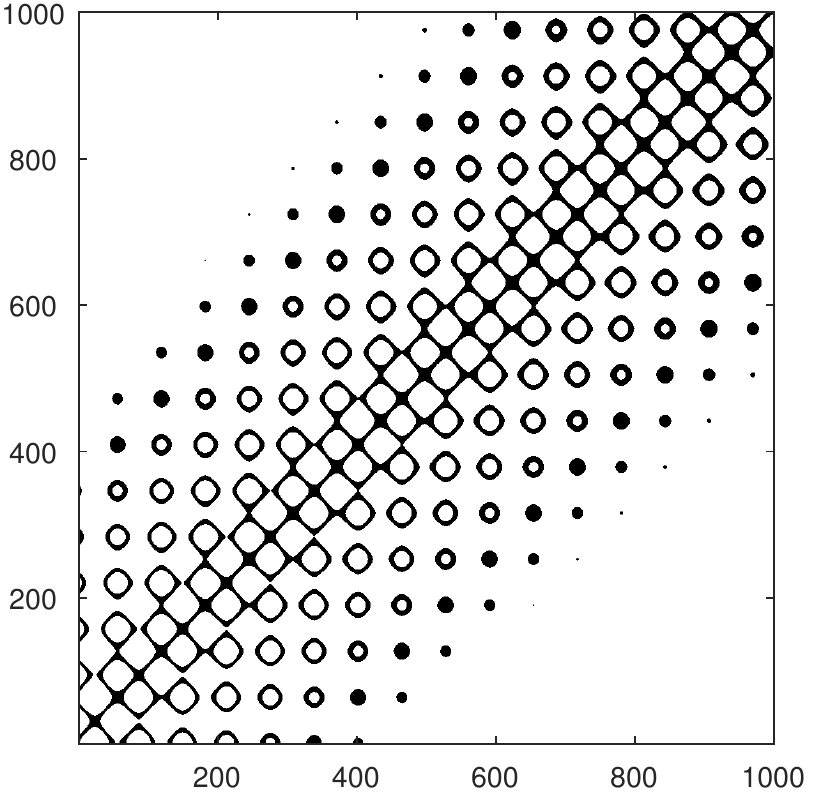}
    \label{fig:sin_drift}}
				
\caption{RPs generated by state sequences $\mathbf{h}$ of ESN fed with input signals associated with the considered classes. Both axes represent time.}
\label{fig:RP_examples}
\end{figure}

We refer to examples in order to observe the different behaviours shown by an RP in correspondence with the aforementioned classes.
We recall that RPs offer visual information and the validity of the following comments are general and not application-specific.
In addition, we stress that RP-based analyses are applicable also to short time-series (with tens/hundreds of records).\\

\noindent\textit{\textbf{Time-dependency}}\\
A uniformly distributed RP is a clear sign of the absence of a time-dependency in the time-series (e.g., an uncorrelated signal).
It is possible to rely on specific RQA measures in order to numerically investigate the presence of time-dependency.
A signature for this is observed by checking the outcomes of DET (\ref{eq:DET}), ENTR (\ref{eq:ENTR}), and SWRP (\ref{eq:swrp}).
All three indexes would yield very low values (close to zero) when there is no time-dependence in the signal.
A periodic signal offers a counter-example having instead a strong time-dependency; DET would be very high for a periodic signal, yet ENTR and SWRP would still be low.
In fact, ENTR and SWRP are conceived to highlight the complexity aspects: signals with low complexity include those with no temporal structure.
As an example, in Fig. \ref{fig:WN_SR01_T04} we examine the RP generated by feeding the ESN with white Gaussian noise, a typical example of signal with no time-dependency.
We observe a uniform RP for the reservoir states, which is peculiar for all signals composed by realizations of statistically independent variables.\\

\noindent\textit{\textbf{Periodicity}}\\
Every periodic system would induce long diagonal lines and the vertical spacing provides the characteristic period of the oscillation.
A periodic system is typically accompanied by high values for DET (\ref{eq:DET}) and $\mathrm{L_{max}}$ (\ref{eq:lmax}). In addition, as stressed before, it has low complexity as expressed by ENTR and SWRP.
In Fig. \ref{fig:LM3.83_SR09_T03}, we show an example of periodic motion generated with a sinusoid having a single dominating frequency. The regularity of the diagonal lines can be immediately recognized from the figure.\\

\noindent\textit{\textbf{Laminarity}}\\
A system presents laminar phases if its state does not change or change very slowly over a number of successive time-steps.
Laminar phases can be visually recognized in an RP by the presence of (fairly) large black rectangles.
Every system possessing laminar phases is characterized by high values for LAM (\ref{eq:lam}).
In order to provide an example (but results are  independently valid), we exploit the logistic map (LM),
\begin{equation}
\label{eq:LM}
\mathbf{x}[n+1] = \tau_{\mathrm{LM}}\mathbf{x}[n](1-\mathbf{x}[n]),
\end{equation}
where usually $\tau_{\mathrm{LM}}\in(0,4]$; here LM (\ref{eq:LM}) is used with initial condition $\mathbf{x}[0] = 0.5$.
In Fig. \ref{fig:LM3.679_SR09_T05} we show the RP when $\tau_{\mathrm{LM}} = 3.679$. The system exhibits chaos-chaos transitions in that configuration. In fact, the related RP is compatible with the one of a (mildly) chaotic system, showing also the presence of laminar phases visualized as large black rectangles.\\

\noindent\textit{\textbf{Chaoticity}}\\
RPs offer a particularly useful visual tool in the case of chaotic dynamics. In fact, such dynamics can be recognized by the presence of erratic and very short diagonal lines. As a consequence, RR (\ref{eq:RR}) would be very low. ENTR and SWRP are two measures of complexity that are useful in characterizing also the degree of chaoticity: the higher the measures, the more chaotic/complex the system.
Chaos is characterized by trajectories diverging exponentially fast. This can be quantified with $\mathrm{L_{max}}$ (\ref{eq:lmax}) and DIV (\ref{eq:div}), whose values would be respectively very low and close to one for systems with a high degree of chaoticity.

As an example, we consider a chaotic system obtained through (\ref{eq:LM}) with $\tau_{\mathrm{LM}} = 4$.
The reservoir dynamics, as shown in the RP in Fig. \ref{fig:LM3.99_SR09_T03}, denotes fully developed chaos, as indicated by the presence of short and erratic diagonal lines only.\\

\noindent\textit{\textbf{Non-stationarity}}\\
Peculiar line patterns observed for all nonstationary signals include large white areas with irregular patterns denoting abrupt changes in the dynamics.
Drift is another typical form of nonstationarity, which is visually recognized in an RP by the fading of recurrences in the upper-left and lower-right corners.
In Fig. \ref{fig:fBmP_SR0.9_T02}, we show an example by feeding the ESN with a well-known nonstationary signal: Brownian motion, a random walk resulting in a nonstationary stochastic process; whose increments correspond to Gaussian white noise, a stationary process. In Fig. \ref{fig:sin_drift}, we show an example of drift by adding a linear trend to a sinusoid.
Nonstationarity can be numerically detected by considering an RQA measure called TREND (not used in our study) and by analyzing the variation of RQA measures when time-delay is applied to the signal (see \cite{marwan2007recurrence} for technical details).

%%%%%%%%%%%%%%%%%%%% EXPERIMENTS %%%%%%%%%%%%%%%%%%%%
\section{Experiments}
\label{sec:exp}

In this section, we consider two time-series generated respectively by an oscillatory and by the Mackey-Glass (MG) dynamical system. We chose these two signals since both of them are often considered as benchmarks for prediction in the ESN literature \cite{Steil2005, verstraeten2007experimental, jaeger2001echo} and they exemplify a very regular and a mildly chaotic system, respectively. 
We perform two experiments. In the first one -- Sec. \ref{sec:exp1} -- we use RPs to visualize the dynamics of reservoirs when driven by a given input signal. When the reservoir operates in a stable regime, RPs of reservoir and input show similar line patterns.

In the second experiment -- Sec. \ref{sec:exp2} -- we use RPs as an analysis tool to tune two critical hyper-parameters, the spectral radius ($\rho$) and the input scaling ($\omega_i$), in an ESN driven by input signal. We study the derived RQA measures for different settings of $\rho$ and $\omega_i$, with the aim to (i) assess the network stability and (ii) determine a suitable setting for these two parameters, where the predictive power of the network is maximized.
Note that we focus on the prediction of the time-series generated by MG system \cite{verstraeten2009quantification}, rather than on the test for learning its chaotic attractor from data \cite{jaeger2001echo}. The latter test requires to enable the output feedback, which leads to an internal dynamic and to a Jacobian matrix, different from the ones considered in this work.

In both experiments, we consider an ESN with no output feedback ($\omega_o = 0$), configured with a standard setting: uniformly distributed weights in $[-1, 1]$ for $\mathrm{W}_i^r$ and $\mathrm{W}_r^r$, percentage of non-zero connections in $\mathrm{W}_r^r$ of 25\%. The readout is trained by setting the regularization parameter in the linear regression to $0.1$. According to the standard drop-out procedure, we discarded the first 100 elements of $\mathbf{h}$ in order to get rid of the ESN transient states. In the first experiment, we set the number of reservoir neurons to $N_r = 50$. In preliminary tests, we noted that the number of reservoir neurons, even if very small (two neurons), does not affect the results obtained for the analysis of the recurrences. Therefore, we do not provide here a study in these terms.
We used the Manhattan distance for evaluating the dissimilarity in the phase space,
\begin{equation}
\label{eq:h_diss}
d(\mathbf{h}[j], \mathbf{h}[i])  = \sum_{n=1}^{N_r} | h_n[j] - h_n[i] |.
\end{equation}
The threshold $\tau_{\mathrm{RP}}$ has been calculated by using a percentage of the average dissimilarity value between the states in $\mathbf{h}$. For simplifying the notation, we directly indicate the percentage referring to $\tau_{\mathrm{RP}}$, e.g., $\tau_{\mathrm{RP}}=0.1$ indicates 10\% of the average distance between all states.
In few cases, to better visualize the RPs, we increase $\tau_{\mathrm{RP}}$ beyond values typically used in the literature.
However, as stated before, our main aim here is to show the agreement between RQA measures and the stability--instability transition together with the corresponding accuracy of the network, rather than providing an \textit{exact} quantification of its characteristics.

Our results are easily reproducible by using the ESN\footnote{http://www.reservoir-computing.org/node/129} and RP\footnote{http://www.recurrence-plot.tk/} toolboxes available online.

\subsection{Visualization of ESN dynamics}
\label{sec:exp1}

% ------- sin -------
The first signal is a sinusoid defined as $\mathbf{x}[k]= \sin(\psi k)$, where $k=1, 2, ..., 5000$ and $\psi = 3/50$.

% SIN RP
\begin{figure}[h!]
\centering

	\subfigure[RP of the input signal]{
	\includegraphics[width=0.4\textwidth]{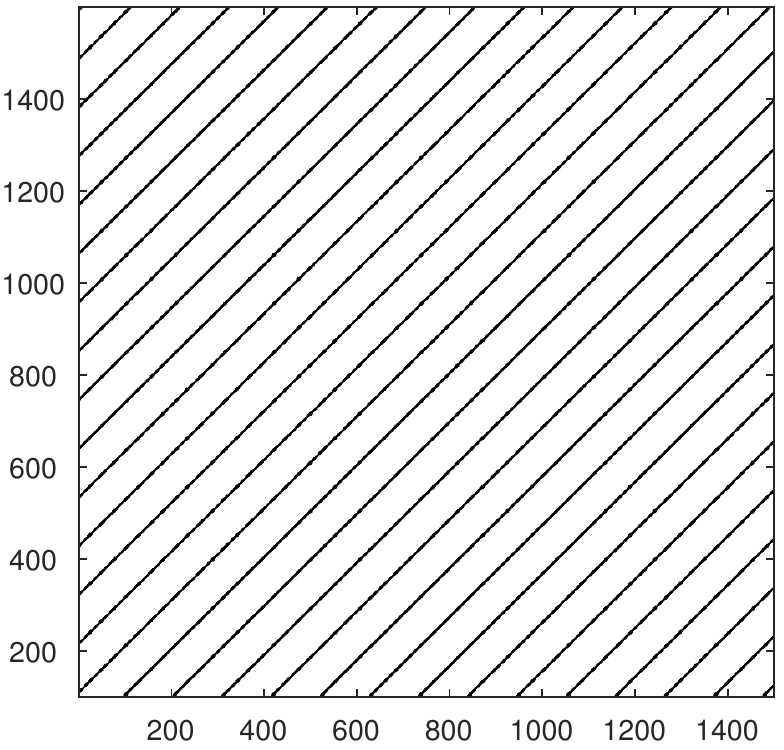}
	\label{fig:sin_input_T02}}\hspace{-1.2em}
	~
	\subfigure[$\rho = 0.99$, $\tau_{\mathrm{RP}} = 0.1$]{
	\includegraphics[width=0.4\textwidth]{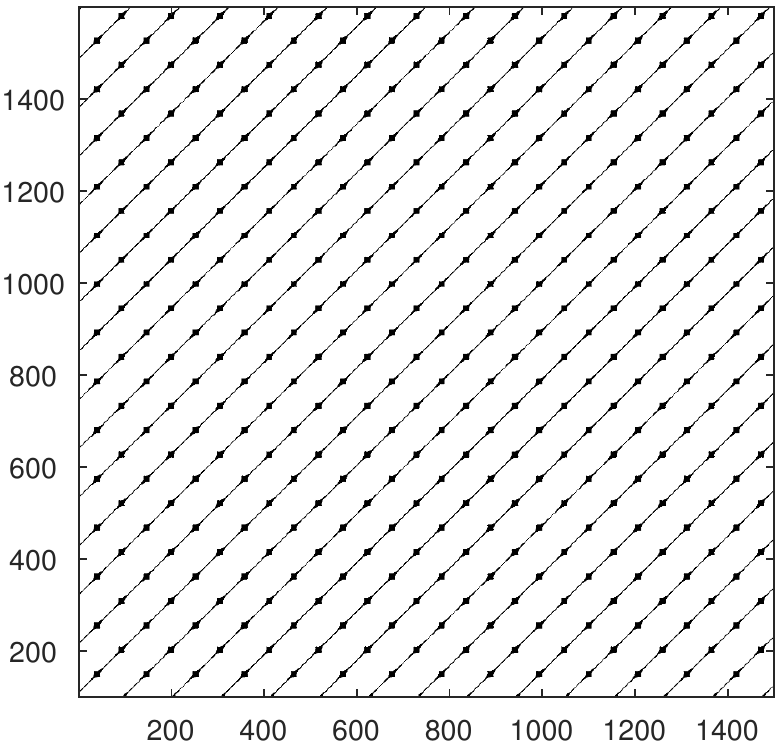}
	\label{fig:sin_SR099_T01}}
				
	\subfigure[$\rho = 1.5$, $\tau_{\mathrm{RP}} = 0.8$]{
	\includegraphics[width=0.4\textwidth]{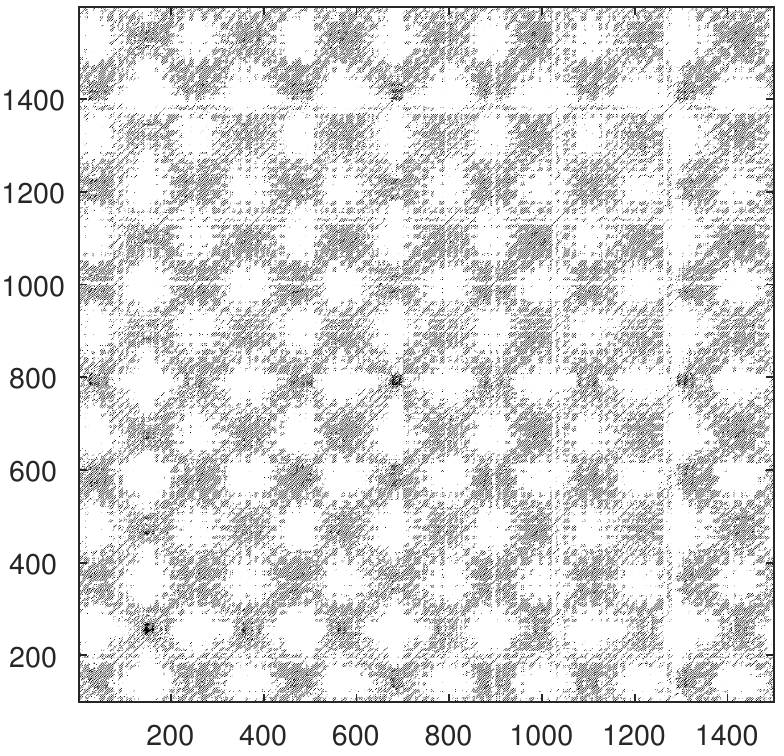}
	\label{fig:sin_SR15_T08}}\hspace{-1.2em}
	~
	\subfigure[$\rho = 2$, $\tau_{\mathrm{RP}} = 0.8$]{
	\includegraphics[width=0.4\textwidth]{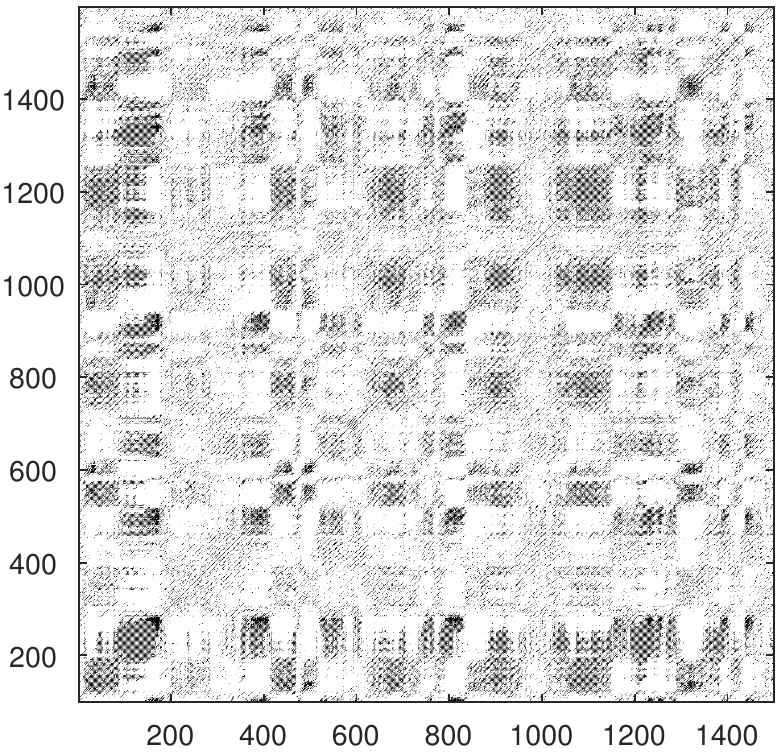}
	\label{fig:sin_SR2_T08}}
				
\caption{RP for the input signal and the sequence of states of the reservoir. When $\rho=0.99$ the activations are compatible with the input dynamics. When $\rho$ exceeds one, the activations denote instability.}
\label{fig:sin}
\end{figure}
In Fig. \ref{fig:sin} we show the RPs relative to the input signal and the reservoir states; input signal has been embedded into a 2-dimensional phase space.
In Fig. \ref{fig:sin_SR099_T01} we used a conservative setting for $\rho$. The resulting RP is compatible with a periodic system having a single frequency. This suggests that, although the (randomly initialized) reservoir performs a non-linear mapping, its resulting dynamics preserves the one of the input system.
In fact, the vertical spacing between the diagonal lines is almost the same, which is 98 in Fig. \ref{fig:sin_input_T02} and 101 in Fig. \ref{fig:sin_SR099_T01}.
This small difference is due to a slight discrepancy in the period of the two systems (i.e., input and reservoir), whose cause was formally explained in \cite{manjunath2013echo}. There the authors demonstrate that when an ESN is driven by a periodic signal with period $\vartheta$, the induced network dynamics asymptotically becomes periodic as well, with period $r \vartheta, r>0$. 
On the other hand, when $\rho$ is pushed beyond one, the RPs are comparable with the ones of unstable/chaotic systems with a degree of instability monotonically related to the value of $\rho$. Please note that, while in Figs. \ref{fig:sin_input_T02} and \ref{fig:sin_SR099_T01} we used $\tau_{\mathrm{RP}} = 0.2$ and $\tau_{\mathrm{RP}} = 0.1$, respectively, in Figs. \ref{fig:sin_SR15_T08} and \ref{fig:sin_SR2_T08} we used a large threshold, $\tau_{\mathrm{RP}} = 0.8$, to improve readability of the plots.

\begin{figure}[ht!]
\centering

    \subfigure[RP of the input signal]{
    \includegraphics[width=0.4\textwidth]{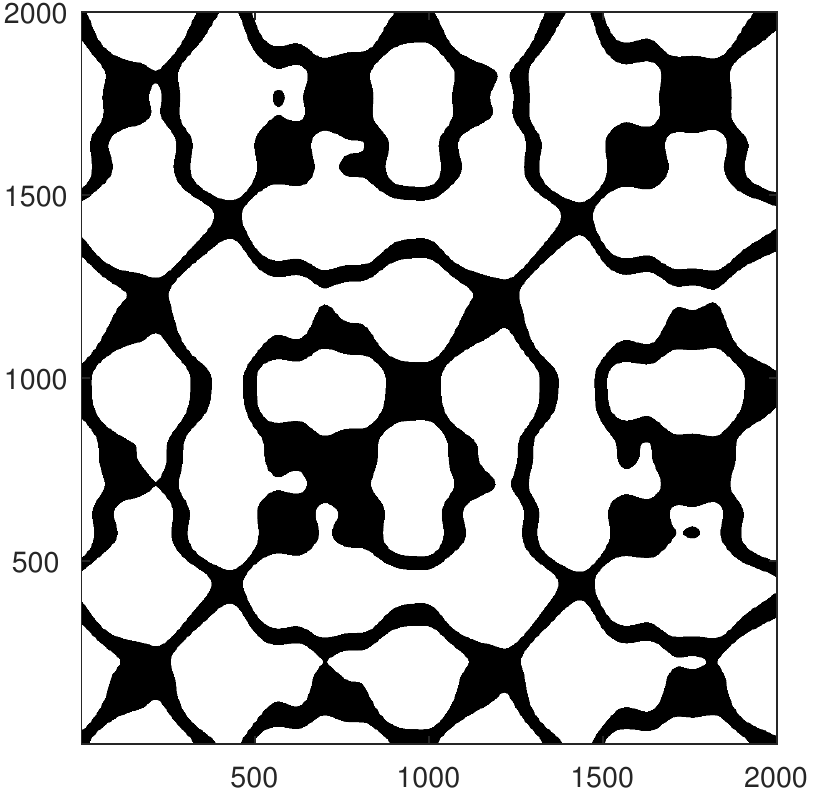}
    \label{fig:MG_input_t05}}\hspace{-1.2em}
    ~
	\subfigure[$\rho = 0.9$, $\tau_{\mathrm{RP}} = 0.5$]{
    \includegraphics[width=0.4\textwidth]{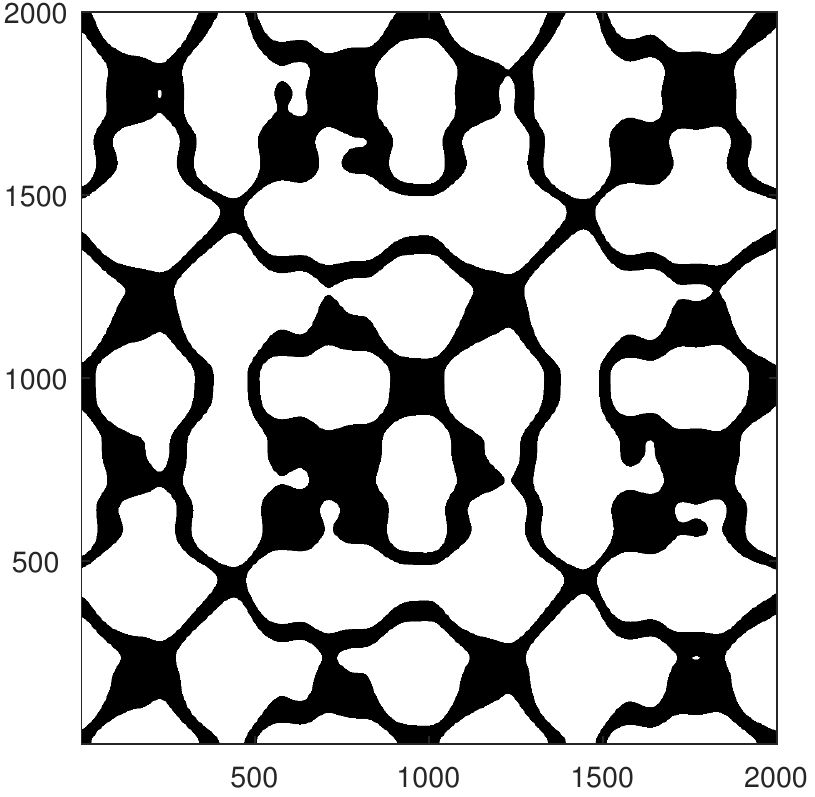}
    \label{fig:MG_SR09_t05}}
		
    \subfigure[$\rho  = 1.5$, $\tau_{\mathrm{RP}} = 0.5$]{
    \includegraphics[width=0.4\textwidth]{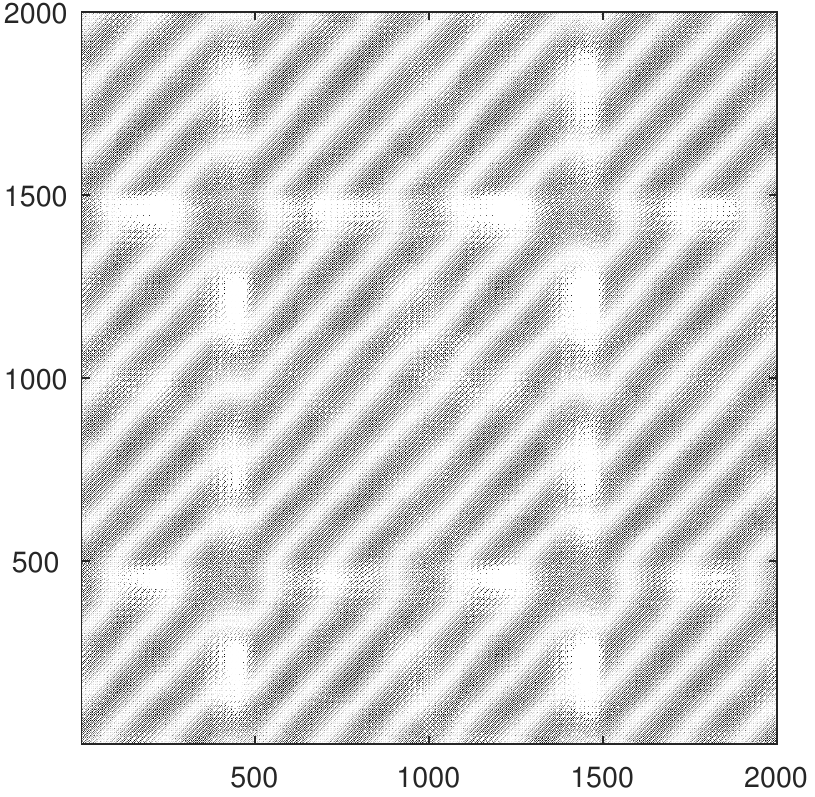}
    \label{fig:MG_SR15_t05}}\hspace{-1.2em}
    ~
    \subfigure[$\rho  = 2$, $\tau_{\mathrm{RP}} = 0.5$]{
    \includegraphics[width=0.4\textwidth]{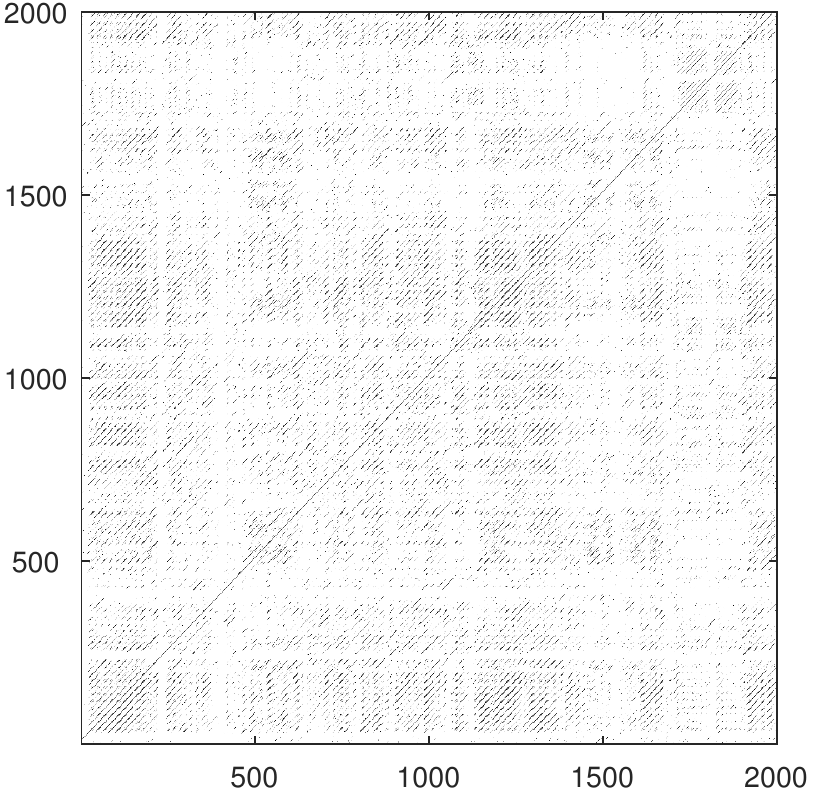}
    \label{fig:MG_SR2_t05}}
		
\caption{RPs of the MG time-series and of the state of the reservoir.}
\label{fig:mg}
\end{figure}

% ------- MG -------
The second input signal is given by a time-series generated from the Mackey-Glass (MG) system,
\begin{equation}
\label{eq:MG}
\frac{dx}{dt} = \frac{\alpha x(t-\tau_{\mathrm{MG}})}{1+ x(t-\tau_{\mathrm{MG}})^{10}} - \beta x(t).
\end{equation}
We obtained a time-series of 150000 time-steps using $\tau_{\mathrm{MG}} = 17, \alpha = 0.2, \beta = 0.1$, initial condition $x(0)=1.2$, and 0.1 as integration step for (\ref{eq:MG}).
In Fig. \ref{fig:mg}, we report RPs of the input signal and those related to the three different settings of the ESN with an increasing value of $\rho$.
As in the previous case, it is worth stressing the similarity between RPs on input in Fig. \ref{fig:MG_input_t05} and reservoir in Fig. \ref{fig:MG_SR09_t05}.
As the reservoir is pushed toward instability (by increasing $\rho$), it is possible to observe the usual incremental transition toward a chaotic regime in the related RPs.
In Figs. \ref{fig:MG_SR15_t05} and \ref{fig:MG_SR2_t05} $\rho$ is set to 1.5 and 2, respectively. In both cases we observe typical line patterns of unstable systems.
However, we note that when $\rho=2$ the system becomes fully chaotic (e.g., see \ref{fig:LM3.99_SR09_T03} for a visual comparison), with very short and erratic diagonal lines.

% MG RP

% ------------------------- CHARACTERIZATION OF RESERVOIR IN TERMS OF PERFORMANCE -------------------------
\subsection{Characterization of reservoirs with RQA}
\label{sec:exp2}

Here we show how RQA measures can be exploited to identify the configuration of the ESN hyper-parameters with maximum computational capability.
We evaluate the accuracy of our methodology by considering sinusoidal and MG time-series.
We compare the results with $\lambda$ (\ref{eq:MLLE}) and the minimal singular value of the Jacobian, $\eta$, whose average value is computed over time; both are widely used in the literature for identifying an optimal network setting in an unsupervised way.
In particular, we consider the regions of the hyper-parameter space where $\lambda$ crosses zero and where $\eta$ is maximized as discriminators to determine the (input-dependent) edge of stability of the network

The error measure for prediction is the Normalized Root Mean Squared Error (NRMSE),
\begin{equation}
\label{eq:nrmse}
\textrm{NRMSE} = \sqrt{\frac{\langle \lVert \mathbf{y} - \mathbf{d} \rVert^2 \rangle}{\langle \lVert \mathbf{y} - \langle\mathbf{d}\rangle \rVert^2 \rangle}},
\end{equation}
being $\mathbf{y}$ the ESN output (\ref{eq:esn_output}) and $\mathbf{d}$ the desired one. The prediction accuracy, $\gamma$, is defined as $\gamma=\max\{0, 1-\mathrm{NRMSE}\}$.

In the following, we monitor how $\gamma$, $\lambda$, $\eta$, and RQA measures vary as the values of $\rho$ and $\omega_i$ are changed.
First, we compare $\lambda$ with $\mathrm{L_{max}}$, the value of the longest diagonal line in an RP -- see Eq. \ref{eq:lmax} and related discussion. $\mathrm{L_{max}}$ is a global indicator of stability that we show here to be highly correlated with $\lambda$.
Successively, we show that the edge of stability determined by using Eq. \ref{eq:edge_criterion} produces results that are significantly better (in a statistical sense) than those derived by using either $\lambda$ or $\eta$.
In order to obtain more interpretable RQA values varying in the unit interval, we normalize all RQA measures by using a unity-based normalization separately for each measure, which considers every setting of $\rho$ and $\omega_i$ for each signal.
It is important to note that this operation neglects the possibility to perform a more in-depth analysis on the RQA measures -- which we remind is not the scope of this paper.
However, this choice allows the network designer to immediately recognize areas of interest and transitions in the related plots.
We always use 50 bins for computing the SWRP measure (\ref{eq:swrp}).
All results are always reported as average of 15 different simulations with independent initialization of $\mathbf{W}_i^r$ and $\mathbf{W}_r^r$ in the ESN.
For this second experiment, since the number of reservoir neurons has an impact on the prediction performance, we adopted two different configurations for $N_r$.\\

% ------------------ Sinusoidal Input ------------------
\textit{\textbf{Sinusoidal signal.}}
The ESN is trained to perform a 25-step ahead prediction and it is tested on a time-series of 1500 time-steps; we used a reservoir size $N_r = 75$.
We test the network performance by varying $\rho \in P = [0.01, 2]$ and $\omega_i \in \Omega = [0.01, 1]$, both discretized with resolution 0.1.

\begin{figure}[!htp]
\centering

				\subfigure[RR]{
        \includegraphics[width=0.26\columnwidth, trim={0.1em 0.1em 0.2em 0.2em},clip]{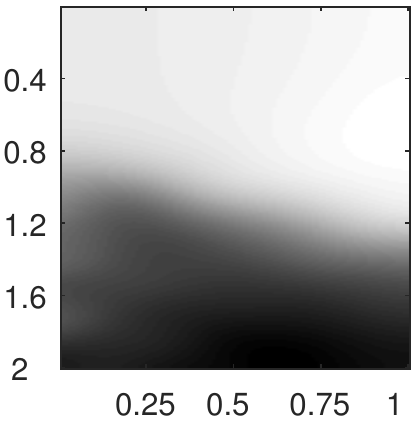}
        }\hspace{-0.5em}%
				~
				\subfigure[DET]{
        \includegraphics[width=0.26\columnwidth, trim={0.1em 0.1em 0.2em 0.2em},clip]{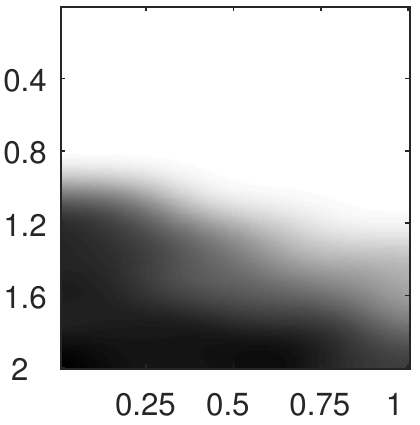}
				}\hspace{-0.5em}%
				~
				\subfigure[$\mathrm{L_{max}} \;\;\;\;\;\;\;\;\;$]{
				\includegraphics[width=0.365\columnwidth, trim={0.1em 0.1em 1.9em 0.9em},clip]{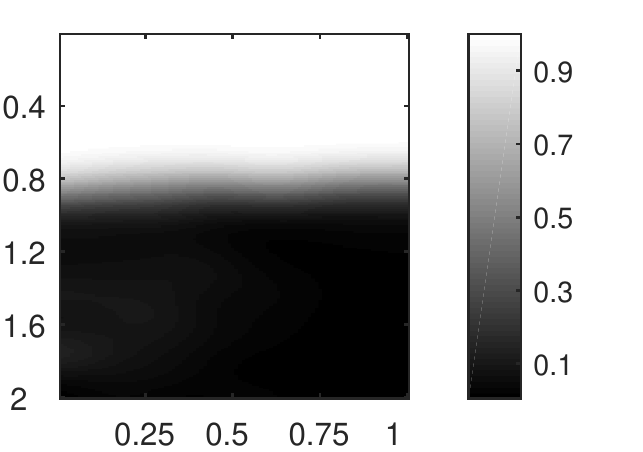}
        }
	
				\subfigure[LAM]{
        \includegraphics[width=0.26\columnwidth, trim={0.1em 0.1em 0.2em 0.2em},clip]{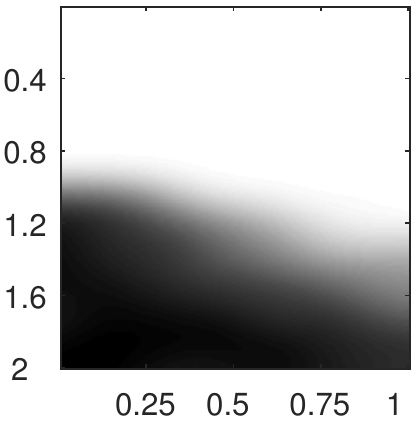}
        }\hspace{-0.5em}%
				~			
				\subfigure[ENTR]{
        \includegraphics[width=0.26\columnwidth, trim={0.1em 0.1em 0.2em 0.2em},clip]{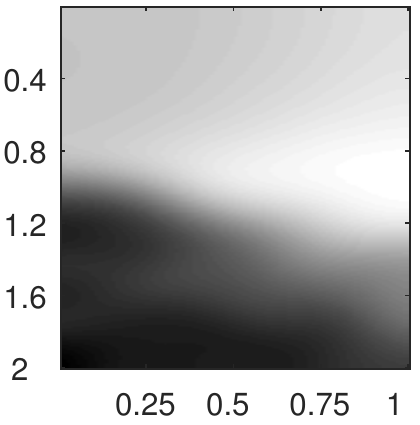}
				}\hspace{-0.5em}%
        ~
				\subfigure[SWRP $\;\;\;\;\;\;\;\;\;$]{
        \includegraphics[width=0.365\columnwidth, trim={0.1em 0.1em 1.9em 0.9em},clip]{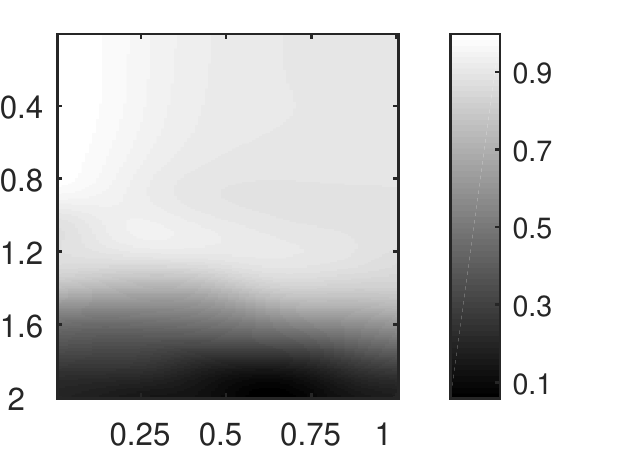}
        }
								
\caption{RQA measures for sinusoid while changing $\rho$ (vertical axis) and $\omega_i$ (horizontal axis).}
\label{fig:sin_RQA}
\end{figure}

In Fig. \ref{fig:sin_RQA}, we show the RQA measures calculated by varying $\rho$ and $\omega_i$.
The ESN, when $\rho \leq 1$, shows high levels of regularity and stability, as expressed by RR, DET, LAM, and $\mathrm{L_{max}}$.
Notably, when $\rho \simeq 1$, we note that $\mathrm{L_{max}}$ starts to decrease, suggesting that we are approaching the unstable phase.
ENTR and SWRP behave similarly, although they are maximized for different configurations of $\rho$ and $\omega_i$.
In fact, ENTR assumes the relative maximum value for larger input scaling when $\rho\simeq 1$. On the other hand, SWRP assumes high values for smaller values of $\omega_i$ and $\rho$.

\begin{figure}[!htp]
\centering
				\subfigure[$\mathrm{L_{max}}$, $\lambda$ and DIV]{
        \includegraphics[width=0.4\textwidth]{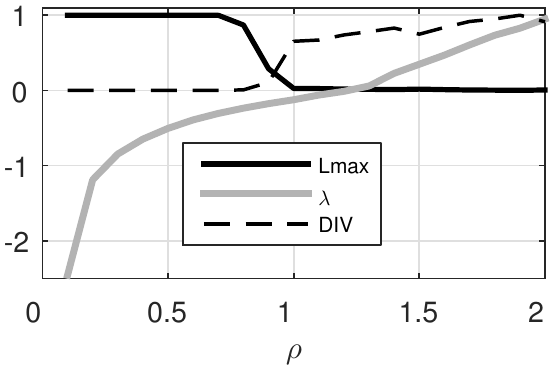}
				\label{fig:sin_MLLE_Lmax_DIV}
        }
				~
				\subfigure[RR]{
        \includegraphics[width=0.4\textwidth]{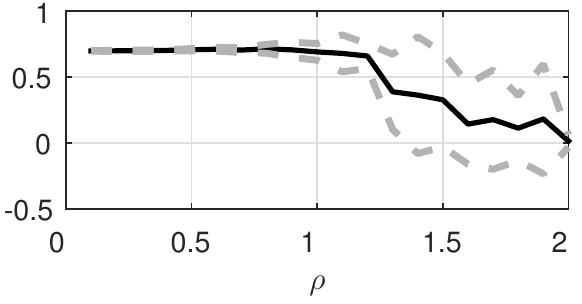}
        }

				\subfigure[DET]{
        \includegraphics[width=0.4\textwidth]{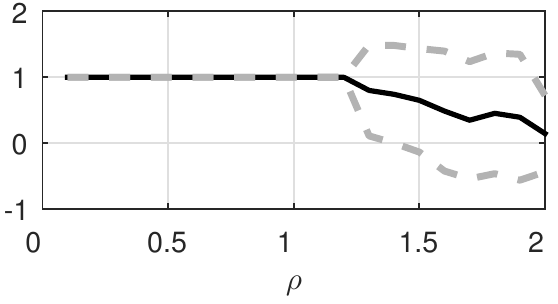}
				}				
				~
				\subfigure[LAM]{
        \includegraphics[width=0.4\textwidth]{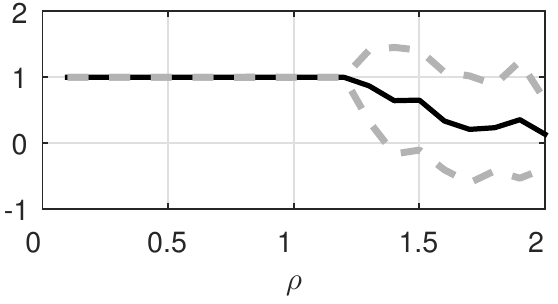}
        }

        \subfigure[ENTR]{
        \includegraphics[width=0.4\textwidth]{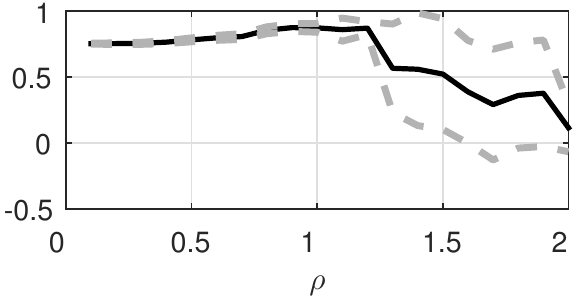}
        }
        ~
				\subfigure[SWRP]{
        \includegraphics[width=0.4\textwidth]{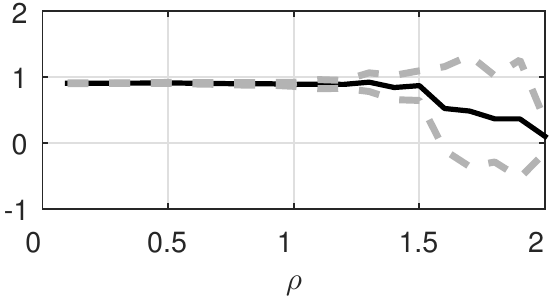}
        }
				
\caption{RQA measures of sinusoidal input for a fixed input scaling, $\omega_i=0.8$. The mean value of each measure is drawn with a solid black line and the standard deviation with gray dashed lines. All RQA measures are stable around their mean values until $\rho$ is pushed beyond 1. Fig. \ref{fig:sin_MLLE_Lmax_DIV} depicts the rescaled, mean value of $\lambda$ (gray solid line), the mean value of $\mathrm{L_{max}}$ (solid black line), and the mean value of DIV (dashed black line).}
\label{fig:sin_RQA_2d}
\end{figure}

\begin{figure}[htp!]
\centering

	\subfigure[$\gamma$]{
	\includegraphics[width=0.26\columnwidth, trim={0.1em 0.1em 1.1em 0.9em},clip]{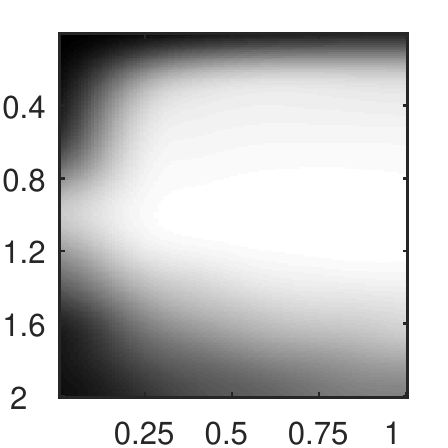}
	\label{fig:sin_NRMSE}}\hspace{-0.5em}%
	~
	\subfigure[$\lambda$]{
	\includegraphics[width=0.26\columnwidth, trim={0.1em 0.1em 1.1em 0.9em},clip]{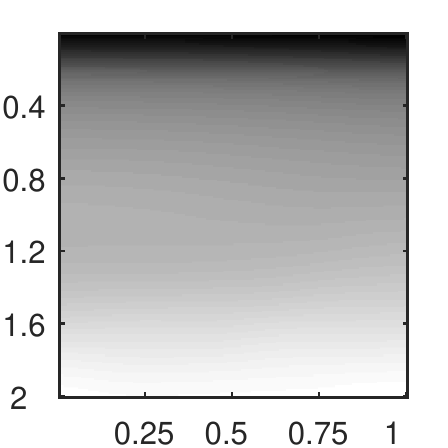}
	\label{fig:sin_MLLE}}\hspace{-0.5em}%
	~
	\subfigure[$\eta$]{
	\includegraphics[width=0.37\columnwidth, trim={0.1em 0.1em 2.1em 0.9em},clip]{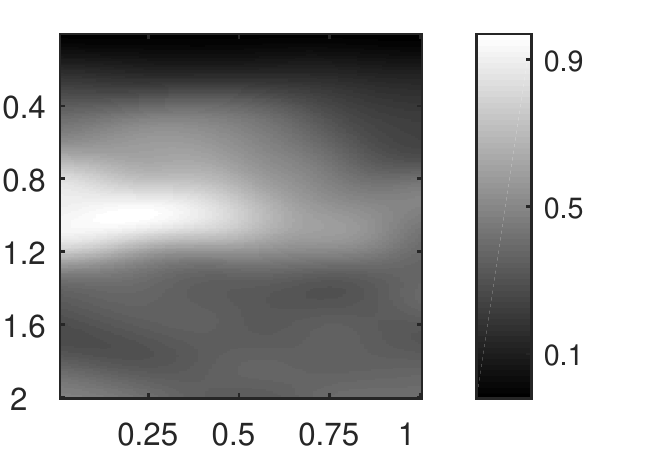}
	\label{fig:sin_SV}}
	
	\subfigure[$\lambda$ (light gray) vs $\mathrm{L_{max}}$ (dark gray)]{
	\includegraphics[width=0.47\columnwidth, trim={1.7em 1.2em 2.1em 2em},clip]{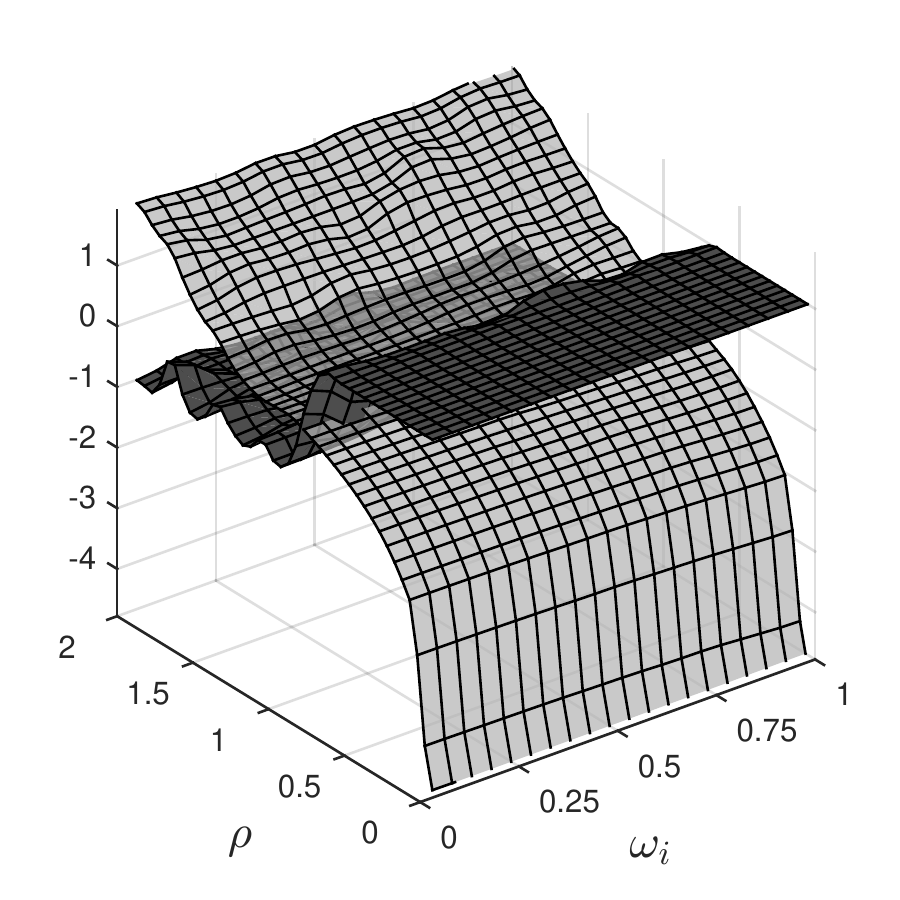}
	\label{fig:sin_MLLE_vs_Lmax}}\hspace{-1em}%
	~		
	\subfigure[Edge of stability]{
	\includegraphics[width=0.47\columnwidth, trim={0.5em 0.4em 2.1em 1.8em},clip]{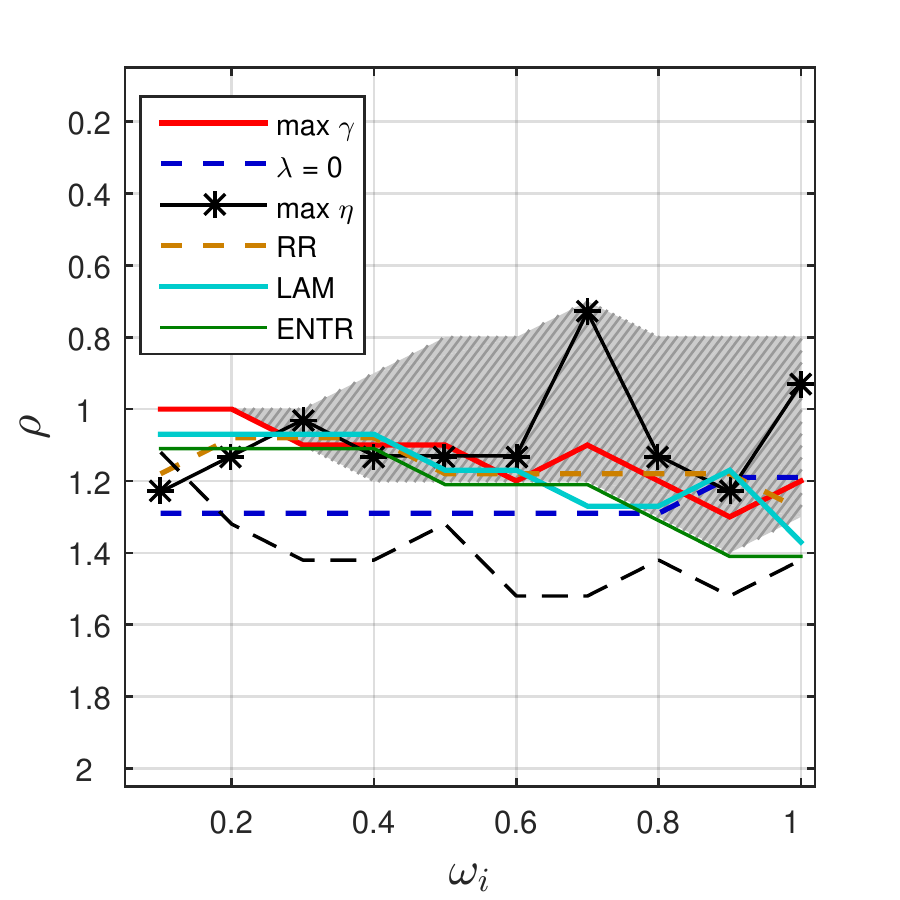}
	\label{fig:sin_Edge}}
	~
			
\caption{Sinusoidal input. 
Prediction accuracy $\gamma$ in Fig. \ref{fig:sin_NRMSE}, $\lambda$ in Fig. \ref{fig:sin_MLLE}, and $\eta$ in Fig. \ref{fig:sin_SV}, calculated for different values of $\rho$ and $\omega_i$.
In Fig. \ref{fig:sin_MLLE_vs_Lmax} $\lambda$ (light gray surface) assumes positive values in correspondence of the gap where $\mathrm{L_{max}}$ (dark gray surface) goes to 0. 
In Fig. \ref{fig:sin_Edge}, the edge of stability is identified as the configurations where $\lambda$ crosses 0, $\eta$ is maximized, and with RQA measures according to Eq. \ref{eq:edge_criterion}). Please notice that, for the sake of readability, here we show only three RQA measures; see Tab. \ref{tab:distancesEdge} for detailed results.
The gray area in Fig. \ref{fig:sin_Edge} shows configurations with high accuracy and it corresponds to the white area in Fig. \ref{fig:sin_NRMSE} where $\gamma$ assumes larger values; the red line where $\gamma$ is maximum.}
\label{fig:sin_performance}
\end{figure}

Fig. \ref{fig:sin_RQA_2d} shows a 2D section of Fig. \ref{fig:sin_RQA}, obtained by selecting a specific input scaling $\omega_i = 0.8$.
In this way, we can visually assess the agreement between $\mathrm{L_{max}}$ and $\lambda$, and also the increasing RQA variability (i.e., standard deviation) on the edge of stability.
In Fig. \ref{fig:sin_MLLE_Lmax_DIV}, we reported $\lambda$, $\mathrm{L_{max}}$, and DIV (\ref{eq:div}) as we vary $\rho$. In order to improve the visualization, $\lambda$ has been rescaled by dividing it for its maximum (absolute) value.
$\lambda$ and $\mathrm{L_{max}}$ are anticorrelated with (Pearson) correlation equal to $-0.78$: the value of $\mathrm{L_{max}}$ decreases as $\rho$ increases, while $\lambda$, as expected, increases with $\rho$.
Additionally, we can observe that there exists a positive correlation ($0.79$) between $\lambda$ and DIV.
We stress that the agreement between $\lambda$ and $\mathrm{L_{max}}$ is consistent for the entire range of $\omega_i$, as represented in Fig. \ref{fig:sin_MLLE_vs_Lmax} and quantified in Tab. \ref{tab:corrStability}, confirming that statistics of the RP diagonal lines offer consistent and solid complexity measures characterizing the network stability.
Please, notice that the correlations in Tab. \ref{tab:corrStability} are computed on two matrices in $\mathbb{R}^{|P|} \times \mathbb{R}^{|\Omega|}$. Since the matrix elements are not interrelated, such matrices can be represented as vectors having dimension $|P|\cdot|\Omega|$. The correlation is then computed on such vectors.  

Let us comment the results for RR, DET, LAM, ENTR, and SWRP with respect to determination of the edge of stability.
All these RQA measures denote low fluctuations until $\lambda$ becomes positive -- see related panels in Fig. \ref{fig:sin_RQA_2d} -- which is achieved for $\rho\simeq1.2$.
$\gamma$s are shown in Fig. \ref{fig:sin_NRMSE}, while in Fig. \ref{fig:sin_Edge} we present different lines in the $\rho$--$\omega_i$ plane, each one denoting a specific determination of the edge of stability (by using either $\lambda$ and the RQA measures). In the same plot, with a solid red line we indicate the configurations where the best performance is achieved.
First, we observe that, in general, for low values of $\omega_i$ we obtain a slightly inferior accuracy. In addition, the gray area in Fig. \ref{fig:sin_NRMSE} is larger for higher values of $\omega_i$ and it includes also several configurations with $\rho=1.2$. This is justified by the fact that high-amplitude signals tend to saturate the nonlinear activation function and cause the poles to shrink toward the origin. This results in a system with a larger stability margin and a more contractive dynamics, hence justifying the use of larger values of $\rho$.

To quantify the accuracy in identifying the edge of stability, we report in Tab. \ref{tab:distancesEdge} the average distances of the edges determined trough RQA measures with respect to the maximum values of $\gamma$ (solid red line in Fig. \ref{fig:sin_Edge}). Our results show that, in all cases except for SWRP, RQA measures determine a more accurate edge of stability with respect to the one obtainable by relying on $\lambda$ and $\eta$.
Notably, RR produces differences with $\gamma$ that are significantly smaller than those of $\lambda$ and $\eta$ ($p<0.0001$); the same holds for DET, LAM, and ENTR ($p<0.0001$), but not for SWRP, whose outcome is significantly worse than $\lambda$ and $\eta$ ($p<0.0001$). Such results have been obtained by using the \textit{t}-test with 0.05 as significance threshold.
The important differences with respect to $\lambda$ might be related to its low sensitivity on $\omega_i$, as can be noticed by looking at Fig. \ref{fig:sin_MLLE}.
On the other hand, $\eta$ is more sensitive to $\omega_i$ than $\lambda$, as can be noted in Fig. \ref{fig:sin_SV}.
In fact, differences in Tab. \ref{tab:distancesEdge} between $\eta$ and $\lambda$ are statistically significant ($p<0.0075$)

Let us discuss the results in Tab. \ref{tab:corrPrediction} in terms of correlation between RQA measures and $\gamma$.
For the sinusoidal input, correlations of RQA measures with $\gamma$ are not very satisfactory (although all correlations are significantly different from zero), with the only exception of SWRP.
However, we note that the estimated (normalized) mutual information indicates, in general, the presence of relevant nonlinear dependencies.\\

% SIN RQA
\vspace{-0.5cm}

% SIN RQA 2D

% SIN MLLE vs Lmax

% ------------------ Mackey Glass ------------------
\textit{\textbf{MG time-series.}}
The prediction of the MG time-series is a common benchmark where ESNs achieved good performance in terms of prediction accuracy \cite{shi2007support,jaeger2004harnessing}. For this test, we train the ESN to perform a 20-step ahead prediction on a test time-series of 2000 time-steps; we set the size of the reservoir to $N_r = 100$; intervals $P$ and $\Omega$ are defined as before.

\begin{figure}[ht!]
\centering

				\subfigure[RR]{
        \includegraphics[width=0.26\columnwidth, trim={0.1em 0.1em 0.2em 0.2em},clip]{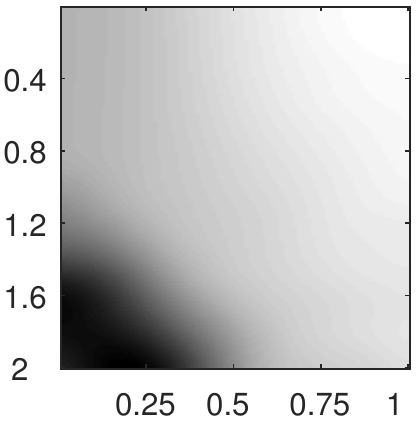}
        }\hspace{-0.5em}%
				~
				\subfigure[DET]{
        \includegraphics[width=0.26\columnwidth, trim={0.1em 0.1em 0.2em 0.2em},clip]{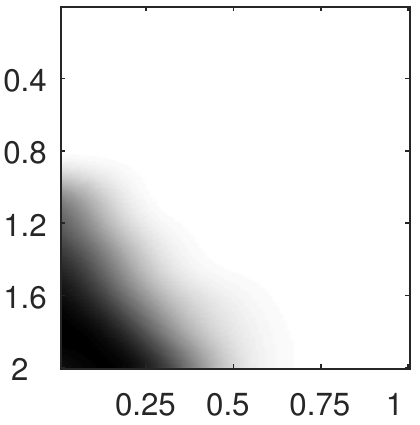}
				}\hspace{-0.5em}%
				~
				\subfigure[$\mathrm{L_{max}}$]{
		    \includegraphics[width=0.365\columnwidth, trim={0.1em 0.1em 1.9em 0.9em},clip]{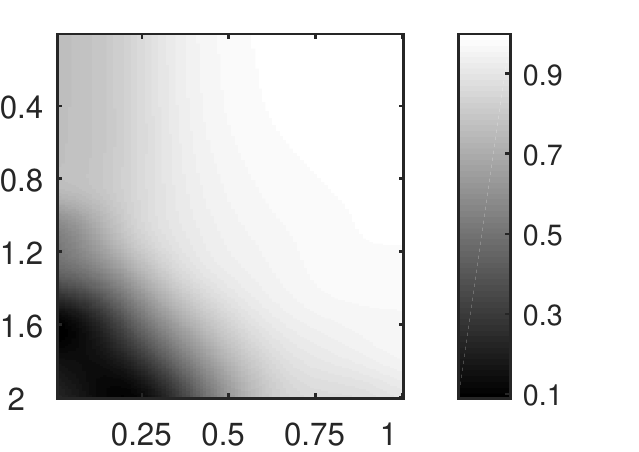}
        }

				\subfigure[LAM]{
        \includegraphics[width=0.26\columnwidth, trim={0.1em 0.1em 0.2em 0.2em},clip]{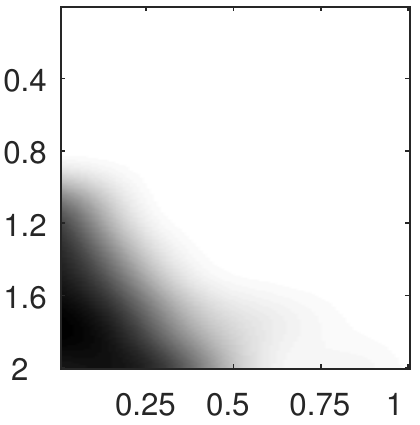}
        }\hspace{-0.5em}%
				~
				\subfigure[ENTR]{
        \includegraphics[width=0.26\columnwidth, trim={0.1em 0.1em 0.2em 0.2em},clip]{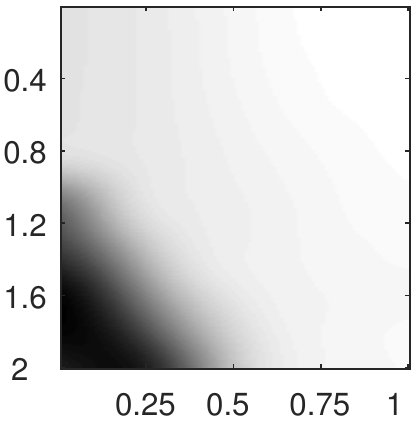}
				}\hspace{-0.5em}%
				~
				\subfigure[SWRP]{
        \includegraphics[width=0.365\columnwidth, trim={0.1em 0.1em 1.9em 0.9em},clip]{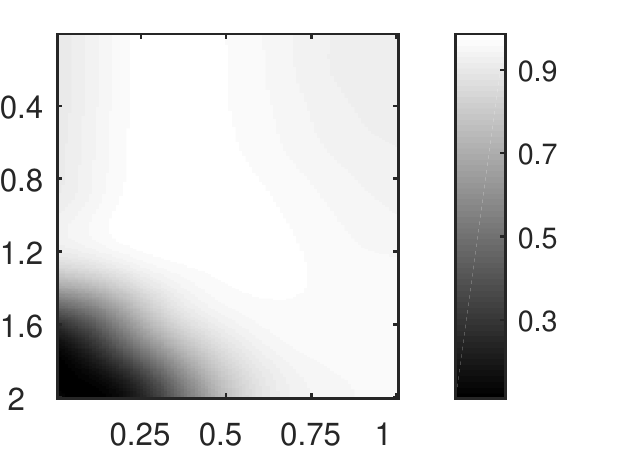}
        }

\caption{RQA measures of MG while changing $\rho$ (vertical axis) and $\omega_i$ (horizontal axis).}
\label{fig:mg_RQA}
\end{figure}

In Fig. \ref{fig:mg_RQA}, we show the outcomes for six RQA measures as both $\rho$ and $\omega_i$ change.
With respect to the sinusoidal input, we immediately observe a higher sensitivity with respect to $\omega_i$.
This can be explained by the need to use a higher degree of non-linearity for dealing with the forecast of the MG time-series.
The dynamics of the reservoir appear as highly deterministic with laminar phases (as expressed by DET and LAM, respectively) for the entire range of $\omega_i$.
We note a consistent maximization for SWRP and ENTR, which assume high values for a large set of configurations that include also values for $\rho\simeq 2$ when $\omega_i > 0.5$.
For what concerns RR, instead, we notice that it is maximized only for very small values of $\rho$ and high values of $\omega_i$, suggesting that density of recurrences (i.e., RR) is very sensitive to $\rho$ for the MG time-series.

\begin{figure}[!ht]
\centering

				\subfigure[$\mathrm{L_{max}}$, $\lambda$ and DIV]{
        \includegraphics[width=0.4\textwidth]{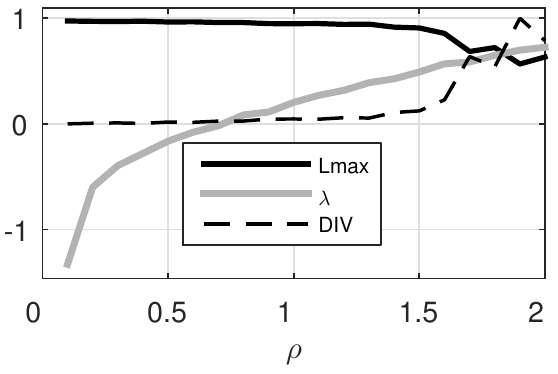}
				\label{fig:mg_MLLE_Lmax_DIV}
        }
				~
				\subfigure[RR]{
        \includegraphics[width=0.4\textwidth]{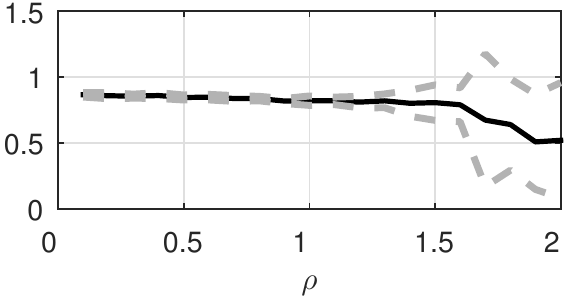}
        }
				
				\subfigure[DET]{
        \includegraphics[width=0.4\textwidth]{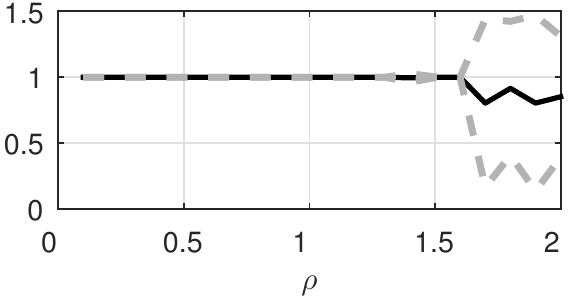}
				}
        ~        
				\subfigure[LAM]{
        \includegraphics[width=0.4\textwidth]{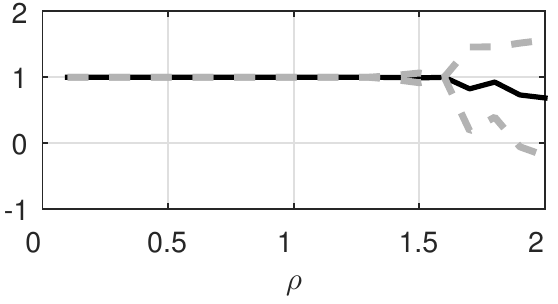}
        }
				
				\subfigure[ENTR]{
        \includegraphics[width=0.4\textwidth]{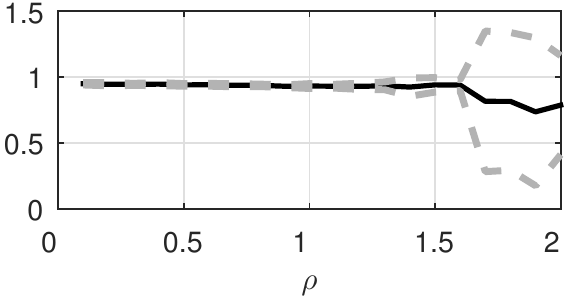}
				}
				~
				\subfigure[SWRP]{
        \includegraphics[width=0.4\textwidth]{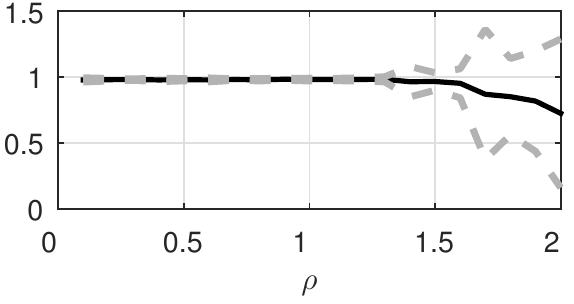}
        }
					
\caption{RQA measures for MG by considering a fixed input scaling, $\omega_i=0.5$. Solid black lines represent the mean value of each measure; gray dashed lines the standard deviation. The standard deviation of every RQA increases significantly as soon as the system enters into unstable regime. In Fig. \ref{fig:mg_MLLE_Lmax_DIV}, we report the rescaled, mean value of $\lambda$ (gray solid line), the mean value of $\mathrm{L_{max}}$ (solid black line), and the mean value of DIV (dashed black line).}
\label{fig:mg_RQA_2d}
\end{figure}

\begin{figure}[htp!]
\centering

    \subfigure[$\gamma$]{
    \includegraphics[width=0.26\columnwidth, trim={0.1em 0.1em 1.1em 0.9em},clip]{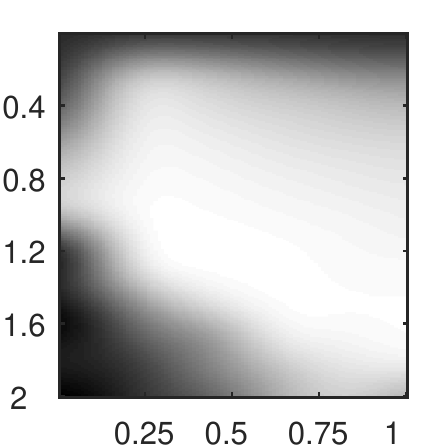}
    \label{fig:mg_NRMSE}}\hspace{-0.5em}%
		~
		\subfigure[$\lambda$]{
    \includegraphics[width=0.26\columnwidth, trim={0.1em 0.1em 1.1em 0.9em},clip]{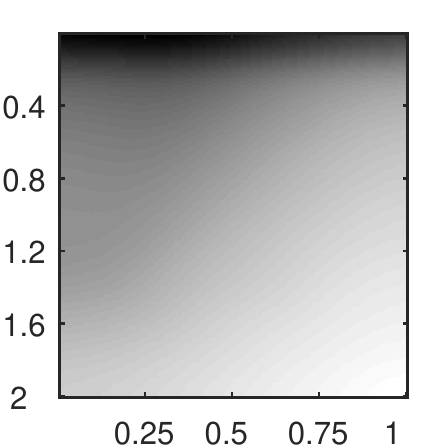}
    \label{fig:mg_MLLE}}\hspace{-0.5em}%
		~
		\subfigure[$\eta$]{
    \includegraphics[width=0.362\columnwidth, trim={0.1em 0.1em 2.1em 0.9em},clip]{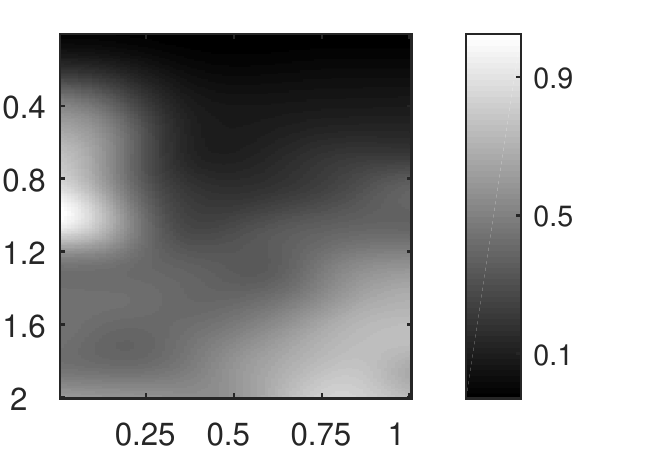}
    \label{fig:mg_minSV}}
    
		\subfigure[$\lambda$ (light gray) vs $\mathrm{L_{max}}$ (dark gray)]{
    \includegraphics[width=0.47\columnwidth, trim={1.7em 1.2em 2em 1.8em},clip]{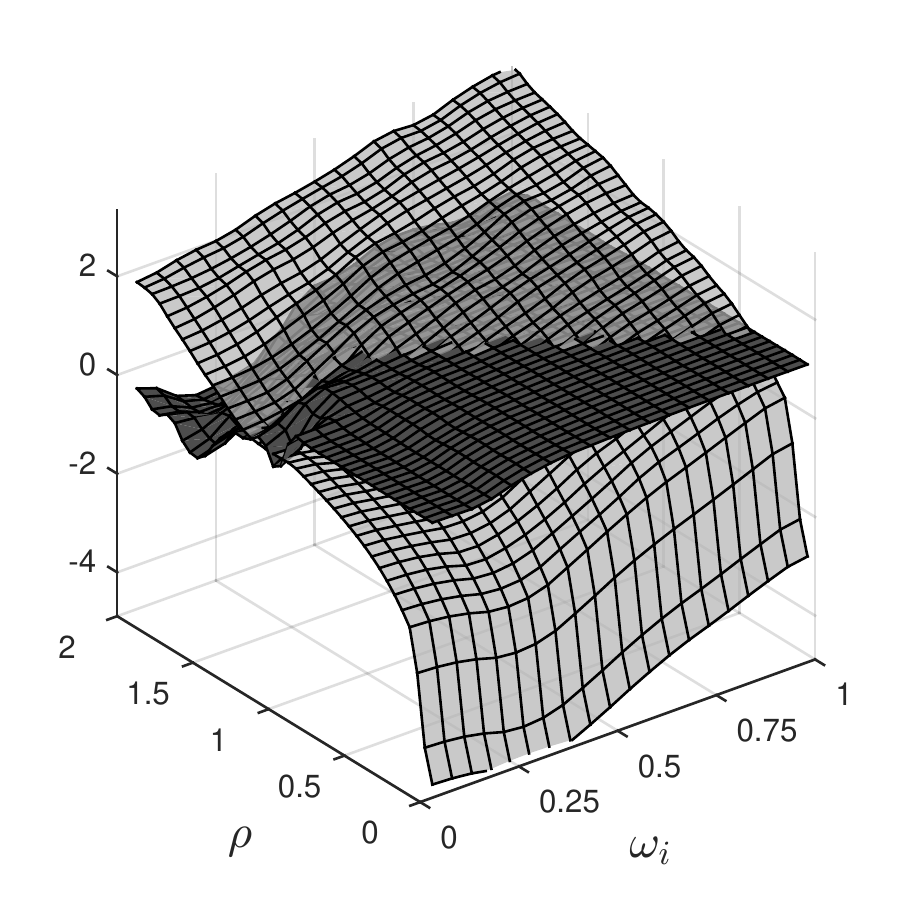}
    \label{fig:mg_MLLE_vs_Lmax}}\hspace{-1em}%
		~  
		\subfigure[Edge of stability]{
		\includegraphics[width=0.47\columnwidth, trim={0.6em 0.2em 2em 1.8em},clip]{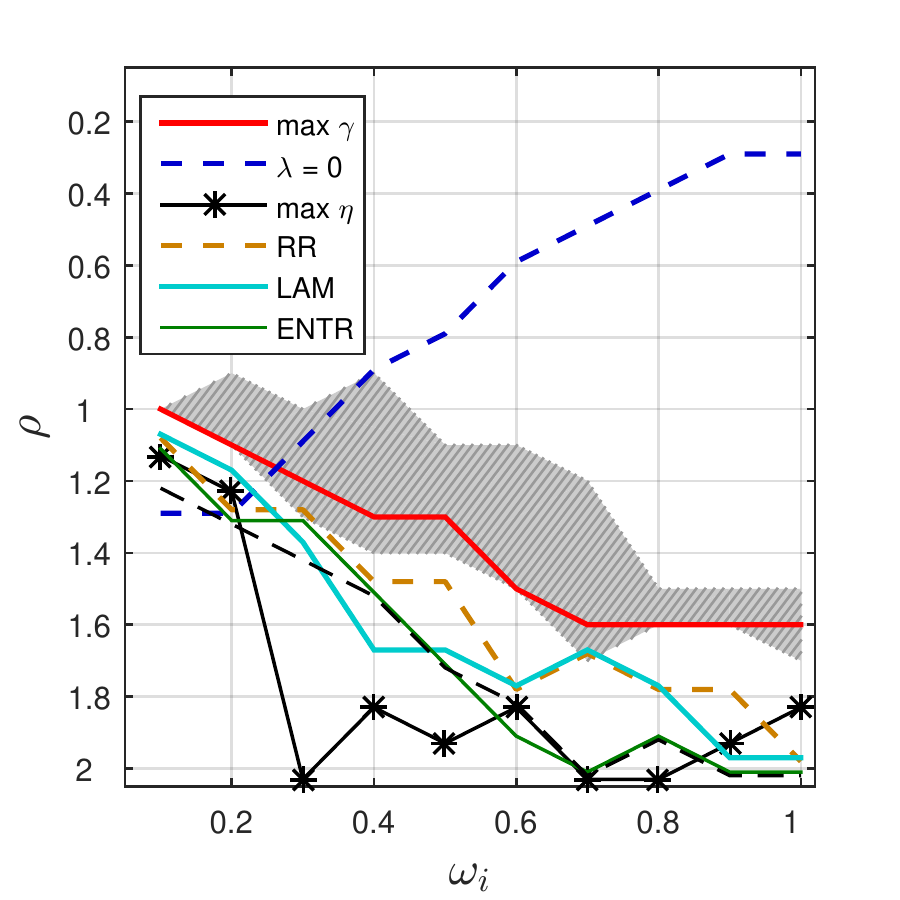}
		\label{fig:mg_Edge}}
		
\caption{MG time-series. 
Prediction accuracy $\gamma$ in Fig. \ref{fig:mg_NRMSE}, $\lambda$ in Fig. \ref{fig:mg_MLLE}, and $\eta$ in Fig. \ref{fig:mg_minSV}, calculated for different values of $\rho$ and $\omega_i$.
In Fig. \ref{fig:mg_MLLE_vs_Lmax} $\lambda$ (light gray surface) becomes positive when the value of $\mathrm{L_{max}}$ (dark gray surface) is drastically reduced to a value close to zero.
In Fig. \ref{fig:mg_Edge}, the edge of stability is identified as the configurations where $\lambda$ crosses 0, $\eta$ is maximized, and for RQA measures according to Eq. \ref{eq:edge_criterion}. Please notice that, for the sake of readability, here we show only three RQA measures; see Tab. \ref{tab:distancesEdge} for detailed results.
The gray area in Fig. \ref{fig:mg_Edge} shows configurations with high accuracy and it corresponds to the white area in Fig. \ref{fig:mg_NRMSE} where $\gamma$ assumes larger values; the red line the values for which $\gamma$ is maximum.}
\end{figure}

In Fig. \ref{fig:mg_RQA_2d}, we show the mean and standard deviation of the RQA values when $\omega_i = 0.5$ is fixed. This is performed, as before, in order to assess the network stability and locate the edge of stability in terms of increase in the fluctuations of the RQA measures.
In Fig. \ref{fig:mg_MLLE_Lmax_DIV} we show the mean value of $\lambda$, which has been graphically rescaled also in this case, $\mathrm{L_{max}}$, and DIV as we vary $\rho$.
As for the sinusoidal input, $\lambda$ and $\mathrm{L_{max}}$ show a good anticorrelation, with a value of $-0.64$. Analogously, $\lambda$ and DIV are correlated with a slightly lower value of $0.60$. The agreement between $\mathrm{L_{max}}$ and $\lambda$ is confirmed for all values of $\omega_i$, as depicted in Fig. \ref{fig:mg_MLLE_vs_Lmax} and numerically quantified in the second row of Tab. \ref{tab:corrStability}.

Let us now take into account the issue of determining the edge of stability.
In Fig. \ref{fig:mg_NRMSE}, we notice that high prediction accuracy $\gamma$ is obtained for configurations located in the center of the plot toward the right-hand side, corresponding to $\rho \simeq 1.5$ and $\omega_i\geq0.8$. For instance, when $\omega_i=0.5$ the best accuracy is obtained for $1\leq\rho\leq1.3$.
When considering the specific case of $\omega_i = 0.5$, $\lambda$ becomes greater than 0 when $\rho\geq0.7$, as shown in Fig. \ref{fig:mg_MLLE_Lmax_DIV}.
However, we note that all RQA measures start to fluctuate for much higher values of $\rho$, close to 1.5.
This is graphically represented in Fig. \ref{fig:mg_Edge}.
As clearly shown by the plot, $\lambda$ significantly underestimates the location of the edge of stability for almost all values of $\omega_i>0.2$.
On the other hand, we note that all RQA measures show a very good correlation with the red line denoting the best performance, $\gamma$, even though they slightly overestimate the edge.
This comparison is made formal in Tab. \ref{tab:distancesEdge}, where we show that all RQA measures denote smaller distances with $\gamma$ than both $\lambda$ and $\eta$, with differences that are statistically significant.
Notably, \textit{p}-values are as follows: for RR, LAM, ENTR, and SWRP $p<0.0001$, and for DET $p<0.0010$.
Therefore, in the case of the MG time-series, the information provided by RQA measures regarding the edge of stability is significantly more accurate than both $\lambda$ and $\eta$.
As in the previous case, we note a low sensitivity of $\lambda$ with respect to the input scaling; see Fig. \ref{fig:mg_MLLE}.
The results obtained for $\eta$ (Fig. \ref{fig:mg_minSV}) show that, instead, this criterion significantly overestimate the edge of stability, especially for values of $\omega_i\in[0.3, 0.5]$.
Finally, it is worth to notice that differences between $\lambda$ and $\eta$ are statistically significant ($p<0.0001$), showing that $\eta$ performs better also in the MG case with respect to $\lambda$.

Let us now take into account the correlations between prediction accuracy and RQA measures.
As shown in Tab. \ref{tab:corrPrediction}, we experience a good linear agreement between SWRP and $\gamma$, denoting a correlation of $0.68$. Also ENTR achieves a good correlation of $0.62$, as in fact ENTR and SWRP behave similarly for the MG time-series (see Fig. \ref{fig:mg_RQA}).
Nonlinear dependency, as indicated by mutual information, shows high values for RR, ENTR, and SWRP.
\bgroup
\def\arraystretch{1.4} %vertical padding
\setlength\tabcolsep{1em} %horizontal padding
\begin{SCtable}[][h!]
\scriptsize
\caption{Correlations between $\lambda$, DIV, and $\mathrm{L_{max}}$.}
\vspace{-0.5cm}
\begin{tabular}{lcc}
\bottomrule
\multicolumn{1}{l}{} & $\mathbf{\lambda} \backslash \mathrm{L_{max}}$ & $\mathbf{\lambda} \backslash \text{DIV}$ \\
\hline
 \textbf{Sin} & -0.74 & 0.53 \\ 
 \textbf{MG} & -0.65 & 0.57 \\
\toprule
\end{tabular}
\label{tab:corrStability}
\end{SCtable}
\egroup

\bgroup
\def\arraystretch{2} %vertical padding
\setlength\tabcolsep{0.3em} %horizontal padding
\begin{table}[th!]\scriptsize
\caption{Average distances and standard deviations between the configurations where $\gamma$ is maximized and the indicators used to determine the edge of stability: $\eta$ is maximum, $\lambda$ crosses 0, and $\sigma_q \geq \bar{\sigma}_q$ for RQA measures. 
Results for RQA shown in bold are statistically significant.
}
\begin{center}
\vspace{-0.3cm}
\begin{tabular}{lccccccc}
\bottomrule
\multicolumn{1}{l}{}  & $\mathbf{\eta}$ & $\mathbf{\lambda}$ & \textbf{RR} & \textbf{DET} & \textbf{LAM} & \textbf{ENTR} & \textbf{SWRP}  \\
\hline
\centering\rotatebox{90}{\hspace{-0.7em} \textbf{Sin}} & 1.4$\pm$1.26& 1.7$\pm$0.95 & \textbf{0.7}$\pm$\textbf{0.67} & \textbf{1.0}$\pm$\textbf{0.82} & \textbf{0.9}$\pm$\textbf{0.74} & \textbf{0.8}$\pm$\textbf{0.63} & 2.5$\pm$0.85 \\ 
\hline
\centering\rotatebox{90}{\hspace{-0.8em} \textbf{MG}}  &3.7$\pm$2.21 & 7.3$\pm$4.78 & \textbf{2.0}$\pm$\textbf{0.94} & \textbf{3.1}$\pm$\textbf{1.29} & \textbf{2.6}$\pm$\textbf{1.35} & \textbf{2.9}$\pm$\textbf{1.29} & \textbf{3.0}$\pm$\textbf{0.94} \\
\toprule
\end{tabular}
\label{tab:distancesEdge}
\end{center}
\end{table}
\egroup

\bgroup
\def\arraystretch{1.3} %vertical padding
\setlength\tabcolsep{1em} %horizontal padding
\begin{table}[th!]\scriptsize
\caption{Dependency between $\gamma$ and RQA measures. First row shows correlations; second row estimated mutual information.}
\begin{center}
\vspace{-0.3cm}
\begin{tabular}{llllll}
\toprule
& \textbf{RR} & \textbf{DET} & \textbf{LAM} & \textbf{ENTR} & \textbf{SWRP}  \\
\hline
\multirow{2}{*}{\centering\rotatebox{90}{\hspace{-0.7em} \textbf{Sin}}} & 0.35 & 0.42 & 0.40 & 0.43 & 0.55 \\
& 0.66 & 0.69 & 0.67 & 0.68 & 0.81 \\
\hline
\multirow{2}{*}{\centering\rotatebox{90}{\hspace{-0.7em} \textbf{MG}}} & 0.57 & 0.40 & 0.38 & 0.62 & 0.68 \\
& 0.77 & 0.55 & 0.54 & 0.78 & 0.80 \\
\bottomrule
\end{tabular}
\label{tab:corrPrediction}
\end{center}
\end{table}
\egroup

% MG RQA

% MG RQA 2D

% MG MLLE vs Lmax

\clearpage
%%%%%%%%%% CONCLUSIONS %%%%%%%%%%
\section{Conclusions}
\label{sec:conclusions}

In this paper, we have investigated the possibility to analyze the dynamics of echo state networks by means of recurrence plots and related RQA complexity measures.
Being a non-autonomous system, its dynamical properties depend not only on the spectral radius of the reservoir matrix, but also on the particular input driving the network and the related scaling coefficients.
We have shown how RPs can be used for reliably analyzing the state evolution of the network when qualitatively different signals are presented as input.
Our results indicates that the recurrences elaborated from the reservoir dynamics clearly denoted an instability level monotonically related to the instability of the network.
This coherent behavior suggests that RPs could be used as a visual tool to design a network.

Our results highlighted a strong statistical agreement between the maximal local Lyapunov exponent and $\mathrm{L_{max}}$, an RQA complexity measure based on the diagonal lines in an RP. This fact could be exploited to define novel, input-dependent stability criteria for reservoirs based on statistics of state recurrences.
We have also examined prediction problems on two signals: a sinusoidal waveform and the Mackey-Glass time-series.
The smooth variation of $\lambda$ was mainly driven by $\rho$ but not very sensitive to the different values used for $\omega_i$.
This fact prevented to accurately identify the input-dependent edge of stability of the network, i.e., the region of the control parameter space where the network departs from a stable regime to reach and operate in an unstable, yet computationally effective phase.
We have considered also a related criterion based on the minimum singular value ($\eta$) of the Jacobian matrix. This criterion offered, in general, better solutions than $\lambda$ for determining the edge of stability. However, both criteria were significantly outperformed by the criterion proposed in this paper based on the fluctuations of RQA measures.
We observed that, when the network approached the edge of stability, all RQA measures abruptly changed their average value, showing also a rapid increase of fluctuations.

The results presented in this paper suggest to allocate future research effort in a number of ways.
First, it would be interesting to study reservoirs initialized according to different (either statistical or deterministic) rules. In addition, it could be interesting to derive a connection between the network short-term memory capacity and RQA measures.
It is worth considering invariant measures elaborated from RPs: the correlation entropy provides a natural lower-bound to the sum of positive Lyapunov exponents.
It might be interesting to study the joint recurrences, calculated between the input system and the reservoir. This might be useful to discover (hidden) mechanisms of synchronization.
Finally, we suggest that such RP-based analyses could be extended to fully trainable recurrent neural networks. The recurrent layer would not be modeled as a multi-dimenional time-series, but instead as a time-varying network. To this end, we hypothesize a possible interplay between graph matching \cite{gm_survey} and recurrence analysis.

\bibliographystyle{IEEEtran}
\bibliography{Bibliography}

\end{document}